\documentclass[usenatbib]{mn2e}
\usepackage{graphicx}
\usepackage[usenames]{color}
\usepackage{euscript}
\usepackage{amssymb}
\usepackage{journals}

\definecolor{grey}{rgb}{0.5,0.6,0.7}

\title[Analytical approximations of $K$-corrections]{Analytical approximations of $K$-corrections in
optical and near-infrared bands}
\author[I. Chilingarian et al.]{Igor V. Chilingarian$^{1,2,3}$\thanks{E-mail:
    Igor.Chilingarian@astro.unistra.fr; chil@sai.msu.ru}, Anne-Laure Melchior$^{1,4}$ and 
    Ivan Yu. Zolotukhin$^{2}$\\
$^{1}$Observatoire de Paris, LERMA, CNRS UMR~8112, 61 Av. de
  l'Observatoire, 75014 Paris, France\\
$^{2}$Sternberg Astronomical Institute, Moscow State University, 13 Universitetskij prospect, 119992, Moscow, Russia\\
$^{3}$Centre de donn\'ees astronomiques de Strasbourg, Observatoire
astronomique de Strasbourg,
UMR~7550, \\\ \ Universit\'e de Strasbourg / CNRS, 11 rue de l'Universit\'e, 67000 Strasbourg, France\\
$^{4}$Universit\'e Pierre et Marie Curie - Paris 6, 4 Place Jussieu, 75252 Paris Cedex 5, France
}
\begin{document}

\date{Accepted 2010 February 10. Received 2010 February 01; in original form 2009 August 14}

\pagerange{\pageref{firstpage}--\pageref{lastpage}} \pubyear{2009}

\maketitle

\label{firstpage}

\begin{abstract} 
To compare photometric properties of galaxies at different redshifts,
the fluxes need to be corrected for the changes of effective rest-frame
wavelengths of filter bandpasses, called $K$-corrections. Usual approaches
to compute them are based on the template fitting of observed spectral
energy distributions (SED) and, thus, require multi-colour photometry. 
Here, we demonstrate that, in cases of widely used optical and near-infrared
filters, $K$-corrections can be precisely approximated as two-dimensional
low-order polynomials of {\em only} two parameters: {\em redshift and one
observed colour}. With this minimalist approach, we present the
polynomial fitting functions for $K$-corrections in SDSS $ugriz$, UKIRT
WFCAM $YJHK$, Johnson-Cousins $UBVR_cI_c$, and 2MASS $JHK_s$ bands for
galaxies at redshifts $Z<0.5$ based on empirically-computed values obtained
by fitting combined optical-NIR SEDs of a set of 10$^5$ galaxies constructed
from SDSS DR7 and UKIDSS DR5 photometry using the Virtual Observatory.
For luminous red galaxies we provide $K$-corrections as functions of their
redshifts only. In two filters, $g$ and $r$, we validate our solutions
by computing $K$-corrections directly from SDSS DR7 spectra. We also present
a $K$-corrections calculator, a web-based service for computing
$K$-corrections on-line.
\end{abstract}


\section{Introduction} 

Extragalactic studies usually require comparison between photometric
data for different galaxy samples, in particular, comparing
measurements obtained for distant galaxies to the local Universe,
where properties of galaxies are studied in a much greater
detail. Generally, any differences in observable parameters arise
from: (1) astrophysical properties of galaxies and (2) observational
biases. The former ones include the galaxy evolution effects: due to
the light travel time, we see distant galaxies as they were
looking several Gyr ago, so the evolution during the last period of
their lifetime simply cannot be observed. On the other hand,
observational biases arise from the process how the observations are
carried out and, therefore, change drastically from one facility to
another, even assuming the data are perfectly reduced and calibrated.
These include effects of aperture, spatial resolution, and a family of
effects connected to the photometric bandpasses.

Photometric data often originate from different observational studies
exploiting different photometric systems and, thus require colour
transformations to be applied (e.g. \citealp{FSI95}). But even if the
data are obtained in the course of one given project, a galaxy sample
may contain objects at different redshifts.

In a broad wavelength range, from ultra-violet to near-infrared, the
spectral energy distributions (SED) of non-active galaxies are mostly
determined by their stellar population properties, i.e. age and
chemical composition, and effects of internal dust attenuation
increasing dramatically at short wavelengths
\citep{CKS94,Fitzpatrick99}.  Stellar population SEDs are very far
from flat distributions and exhibit prominent features
(e.g. \citealp{FR97,BC03a}). At the same time, redshifting the galaxy
spectrum is equivalent to shifting the corresponding filter
transmission curve. This explains the difference in fluxes in the same
bandpass for two hypothetical galaxies having exactly identical SEDs
but being at different redshifts. Historically, this difference is
called $K$-correction \citep{OS68}. The $K$-correction formalism is
presented in detail and thoroughly discussed in \citet{HBBE02,BR07}.

Nowadays, in the era of large wide-area photometric and spectroscopic
extragalactic surveys, the precise, fast and simple computation of
$K$-corrections has become a crucial point for the successful astrophysical
interpretation of data. Several approaches were presented in the literature
\citep{FSI95,Mannucci+01,BR07,RBH09}. \citet{BR07} provide a software
package to compute $K$-corrections for datasets in any photometric
system.  However, since their method is based on the SED fitting
technique, the results critically depend on the availability of
multi-colour photometric data. \citet{FSI95} and \citet{Mannucci+01}
provide only qualitative dependence of $K$-corrections on redshift and
galaxy morphological type; the latter is very difficult to assess in
an automatic way and these methods therefore require availability of
original galaxy images in addition to photometric measurements.

The aim of our study is to explore the parameter space of typical observed
galaxy properties and to provide simple and precise analytical
approximations of $K$-corrections in widely used optical and NIR photometric
bandpasses, based on the minimal set of observables. To achieve this goal,
we exploit a large homogeneous database of optical-to-NIR galaxy SEDs
compiled from modern wide-area photometric surveys. In the next section, we
describe our galaxy sample, details on the computation of $K$-corrections
using {\sc pegase.2} stellar population models \citep{FR97} and comparison
of obtained values with those computed using the {\sc kcorrect} code
\citep{BR07}. In Section~3, we describe analytical approximations, and the
validation of our results using spectral-based $K$-corrections. In
Section~4, we compare our results with the literature and briefly
discuss some astrophysical interpretation of our technique. Appendices
provide tables with coefficients of best-fitting polynomials and
present a ``K-corrections calculator'' service.

\section{Empirical computation of $K$-corrections}

\subsection{Galaxy sample}

We compute $K$-corrections using a large sample of optical-to-NIR SEDs
of nearby galaxies constructed using Virtual Observatory technologies
to retrieve and combine photometric measurements from Sloan Digital
Sky Survey Data Release 7 (SDSS DR7, \citealp{SDSS_DR7}) and the UKIRT
Infrared Deep Survey Data Release 5 (UKIDSS DR5,
\citealp{Lawrence+07}). Comprehensive description of this multi-colour
photometric catalogue will be provided in a separate paper
(Chilingarian et al. in prep), here we give a brief summary essential
for understanding the empirical $K$-correction computations.

We constructed a sample of galaxies excluding broad-line active
galactic nuclei (AGN) by performing the spatial cross-matching of the
SDSS DR7 spectroscopic sample in stripes 9 to 16 with the Large Area
Survey catalogue of UKIDSS DR5. We selected 190,275 galaxies having
spectroscopic redshifts in a range $0.03 < Z < 0.6$ provided by SDSS
DR7 using the SDSS CASJobs
Service\footnote{http://cas.sdss.org/CasJobs}. The spatial
cross-identification with UKIDSS DR5 with a search radius of 3~arcsec
selecting the best positional matches in case of multiple
objects within this radii resulted in 170,533 objects, 87,161 of which
were detected in all four UKIDSS Large Area Survey photometric
bands ($Y$, $J$, $H$, and $K$).

In order to compute $K$-corrections, the photometric measurements from the
two data sources have to be homogeneous. We use SDSS fibre magnitudes
corresponding to $d = 3$~arcsec circular apertures (\emph{fiberMags}), and
computed the corresponding 3~arcsec aperture magnitudes for UKIDSS objects
by linearly interpolating between the values provided for three apertures (2.0, 2.8,
and 5.7~arcsec) applying zero-point corrections \citep{HWLH06} for
converting UKIDSS magnitudes from the Vega into the AB system.

The lower redshift limit, $Z = 0.03$, is selected in order to minimise the
aperture effects: at a distance of 120~Mpc, corresponding to this
redshift, the 3~arcsec aperture encloses about 1.75~kpc, i.e. significant
part of the bulge even for giant galaxies, thus stellar populations in 
galactic nuclei would not dominate the light. Beyond the selected upper
redshift limit, $Z = 0.6$, the fraction of normal galaxies in SDSS
significantly decreases, because of the magnitude-limited selection of 
SDSS spectroscopic targets, and at the same time the quality of absorption-line spectra
becomes quite poor.

We use magnitudes in 3~arcsec apertures and not the Petrosian magnitudes to
be able to compare them directly with the SDSS DR7 spectra obtained within
the same apertures. The median 3~arcsec aperture magnitude uncertainties are
0.01~mag and better for $g$, $r$, $i$, $Y$, $J$, $H$, and $K$, 0.017~mag for
$z$, and 0.07~mag for the $u$ band respectively. All magnitudes are
corrected for the foreground Galactic extinction according to \citet{SFD98}.

\begin{figure}
\includegraphics[width=\hsize]{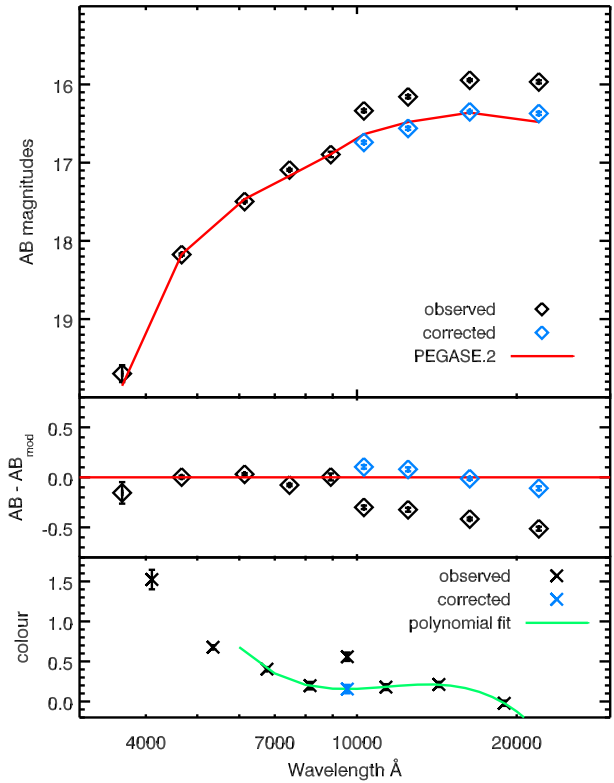}
\caption{Top panel presents the optical-to-NIR SED of
SDSS~J155023.03-000023.8 in a 3-arcsec aperture as provided in the catalogue
(black diamonds), its NIR part empirically corrected (blue diamonds), its
best-matching {\sc pegase.2} template used for $K$-correction computation
(solid red line). Fitting residuals are shown in middle panel (same
symbols). Bottom panel displays colours as a function of wavelength
(black crosses), the best-fitting 3rd order polynomial for 5 colours ($r -
i$, $i - z$, $Y - J$, $J - H$, and $H - K$) is shown as a solid green line;
the corrected value of $z - Y$ is denoted by the blue cross. \label{figSED}}
\end{figure}

There are important systematic offsets of unknown origin between SDSS DR7
fibre magnitudes and UKIDSS DR5 photometry in 3~arcsec apertures.
We fit 3rd order polynomial using 5 colours starting from $r - i$ and
redder except $z - Y$ and compute the offset between the expected $z - Y$
value from the best-fitting polynomial and the observed one. The offset has
a mean value of 0.22~mag and a standard deviation of 0.13~mag independent
from other parameters (e.g. observed colours and a redshift). We therefore
subtract it from all UKIDSS magnitudes. This effect is illustrated in
Fig.~\ref{figSED}. In the top panel, we display the combined SDSS-UKIDSS SED
of some galaxy and its best-matching {\sc pegase.2} template obtained by the
$K$-correction determination procedure described below. The middle and
bottom panels show the residuals between the observed SED and its model, and
colours in consequent spectral bands with the best-fitting 3rd order
polynomial. From the middle panel, it is clear that the NIR part of the SED
is offset by a constant value from the optical part, which is evident as the
measured $z - Y$ colour strongly deviates from the fitting polynomial.

\subsection{Computation of $K$-corrections}

We used two approaches to compute $K$-corrections: (1) the {\sc kcorrect}
software package by \cite{BR07} and (2) {\sc pegase.2}-based computations
described hereafter. The latter technique allows us to roughly estimate mean
stellar population properties as well as internal extinction in galaxies.
The analytical fitting is performed for both methods independently.

We first ran the {\sc kcorrect} software package to compute the
$K$-corrections for our sample of galaxies. This package is based on a
mathematical algorithm, namely non-negative matrix factorisation, which
creates model-based template sets. The initial set of hundreds of templates
is reduced to a basis of five, which in principle can be used to interpret
the galaxy SED in terms of stellar populations. The linear combination of
these templates is then fitted into the set of broadband fluxes available
for each galaxy to derive the $K$-corrections in all bands.

Second, we computed a grid of Simple Stellar Populations (SSP) using the {\sc
pegase.2} evolutionary synthesis code \citep{FR97} for a set of 75 ages
nearly logarithmically spaced between 25~Myr and 16.5~Gyr and 10
metallicities between $-2.5 < [$Fe/H$] < +1.0$~dex. Such a grid was computed
separately for redshifts between $0 < Z < 0.6$ with a step of 0.05. We apply
the \citet{Fitzpatrick99} extinction law for each of 750 SSPs at each
redshift varying the $A_V$ between 0 and 2.25~mag with a step of 0.15~mag
ending up with 11250 template SEDs per redshift.

In order to compute $K$-corrections for a given galaxy, we perform a linear
interpolation of the SSP grid to its redshift, then pick up the
best-matching template SED in terms of $\chi^2$ normalising both data
and templates by the mean fluxes in all filters. Since our photometric
uncertainties are quite small, this approach would not result in significant
biases. Once the best-matching template has been found, the $K$-corrections
in all photometric bands are computed as $K_f(Z) = -2.5 \log (F(0) / F(Z))$,
where $F(Z)$ and $F(0)$ are fluxes in a given filter at redshift $Z$ and in
the restframe. We provide an example of an observed SED and its
best-matching template in the upper panel of Fig~\ref{figSED}.

Since we can calculate model fluxes in any photometric band, we also
used the same technique to compute $K$-corrections in Johnson-Cousins
$UBVR_cI_c$ and 2MASS $JHK_s$ bands.

\begin{figure}
\includegraphics[width=\hsize]{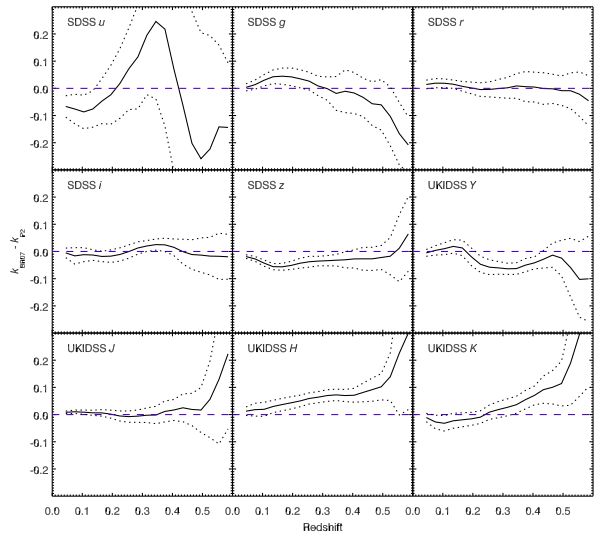}
\caption{Comparison of empirically calculated $K$-corrections obtained by
fitting the photometric data using {\sc pegase.2} SSP models ($k_{P2}$) and by the
{\sc kcorrect} ($k_{BR07}$) code. Each panel displays the difference between 
the two approaches as a function of redshift. Solid lines denote the
median differences and their standard deviations are shown with dotted
lines. \label{figcompKcorr}}
\end{figure}

We compare the values computed in this fashion with the value derived by the
{\sc kcorrect} code from the same photometric dataset. The comparison for
all 9 SDSS-UKIDSS bands is provided in Fig~\ref{figcompKcorr}. 

In general, the results obtained by using the two approaches in
$ugrizYJHK$ bands are quite similar.  However, in certain
spectral bands, some statistically significant differences are
evident. The worst situation is observed in the SDSS $u$ band mainly
for two reasons: relatively poor quality of the $u$ band photometry
especially for objects at higher redshifts and very high sensitivity
of UV colours to even low mass fractions of recently formed stars,
which affect fluxes at longer wavelengths much weaker. Therefore, the
SSP fitting approach presumably should not work well. At the same
time, the linear combination of 5 templates used by the {\sc kcorrect}
package may also produce significant biases due to age and/or
metallicity mismatch between the templates and real galaxies. Finally,
we provide the $K$-corrections for the $u$ band, emphasising that
since no independent verification is possible in our case, one has to
use these results with caution. The same statement applies to $K$ band
$K$-corrections, although the results of the two approaches well match
each other, because the computation relies on the extrapolation of
galaxy SED in the NIR part, where stellar population models and,
correspondingly, the template spectra are of much lower quality than
in the optical wavelength domain.

Computations in the $r$, $i$, and $J$ bands agree remarkably well except the
high-redshift ($Z > 0.5$) end of the $J$ band. There are some systematic
discrepancies between the two techniques in the $g$, $z$, $Y$, and $H$
bands. They may originate from the fact that {\sc pegase.2} SSPs are built
using the theoretical stellar library, shown to introduce colour differences
between synthetic spectra and observed ones at least in the SDSS photometric
system \citep{Maraston+09}. We will analyse the $g$ and $r$ band results below using
direct spectral-based $K$-corrections, while for the remaining bandpasses no
independent test can be performed since no large samples of galaxy spectra
are available in those wavelength domains. However, we note that the
discrepancies are an order of $0.05$~mag, hence, both approaches may be used
in the photometric studies. Therefore, we will proceed with the rest of our
analysis using both techniques of the $K$-correction computation, addressing
them as \textit{BR07} and \textit{SSP} for the {\sc kcorrect} and {\sc
pegase.2} SSP-based approaches respectively.

\section{Results}

\subsection{Analytical approximations}

We observe a large scatter of $K$-corrections as functions of redshift 
reaching 2~mag in
all SDSS-UKIDSS bands except $H$ and $K$. However, exploring the data with the {\sc
topcat} software\footnote{http://www.star.bris.ac.uk/$\sim$mbt/topcat/}, we
found that adding just one observed colour as a second parameter and
approximating the $K$-correction as a surface in the three-dimensional space,
significantly reduces the residual scatter bringing it to the order of
$K$-correction computation uncertainties. We fit $K$-correction values in
every filter $q$ as a polynomial surface of a form:
\begin{equation}
K_{q}(Z, m_{f_1} - m_{f_2}) = \sum_{x=0}^{N_{Z}} \sum_{y=0}^{N_c} 
a_{x,y} Z^{x} (m_{f_1} - m_{f_2})^{y},
\end{equation}
\noindent
where $a_{x,y}$ are polynomial coefficients, $Z$ is a spectroscopic redshift, $m_{f_1}$ and $m_{f_2}$ are observed
magnitudes in filters $f_1$ and $f_2$ chosen for every filter $q$, $N_Z$ and
$N_{c}$ are empirically selected polynomial powers in the redshift and
colour dimensions respectively. Given that $K$-corrections are zero by
definition at $Z=0$, no constant term is needed and all $a_{0,y} = 0$.

Since both $K$-correction computation techniques used in our study are based
on the stellar population models, and our analytical approximations exploit
simple polynomial fitting technique without any clipping of outliers, we
have to exclude strong active galactic nuclei (AGN) and quasars, as well as
SDSS targeting artifacts (e.g. aircrafts, satellites, minor planets) and
objects with wrong redshift determinations, which may affect our best-fitting
solutions. We have fitted all 190,275 SDSS DR7 spectra using the {\sc
nbursts} full spectral fitting technique \citep{CPSA07,CPSK07} in the
restframe wavelength range between 3900 and 6750\AA, and selected for our
further analysis only 164,108 objects having reduced $\chi^2/DOF < 0.9$
(median $\chi^2/DOF = 0.67$, which is smaller than unity because of slight
spectral oversampling of SDSS data). The interpretation of the spectral
fitting results will be provided together with the presentation of a
spectrophotometric catalogue in Chilingarian et al. (in prep).

We computed and fitted $K$-corrections in this fashion for a sample of
164,108 SDSS DR7 galaxies with well-fitted spectra using only SDSS 5-bands
photometry, and to the merged SDSS--UKIDSS sample containing photometric
information in all 9 bands for 74,254 galaxies. While the results from the
two approaches remain nearly statistically identical in $u$, $g$, and $r$
bands, the difference becomes significant in $i$ and especially in $z$. Two
distinct sequences become clear in the redshift vs $K$-correction plots in
$i$ and $z$ in case of the 5-bands based computation. Only one of the two
sequences remains in each case if full 9-band SEDs are used, suggesting that
the second sequence is created by wrong stellar population templates picked
up from the template grid. This clearly demonstrates the importance of the
NIR part to compute $K$-corrections if using a fully empirical approach.
Consequently, r.m.s of the polynomial surface fitting residuals in case of
9-bands based $K$-corrections in the $z$ band is four times smaller compared
to the 5 bands.

\begin{figure*}
\includegraphics[width=0.33\hsize]{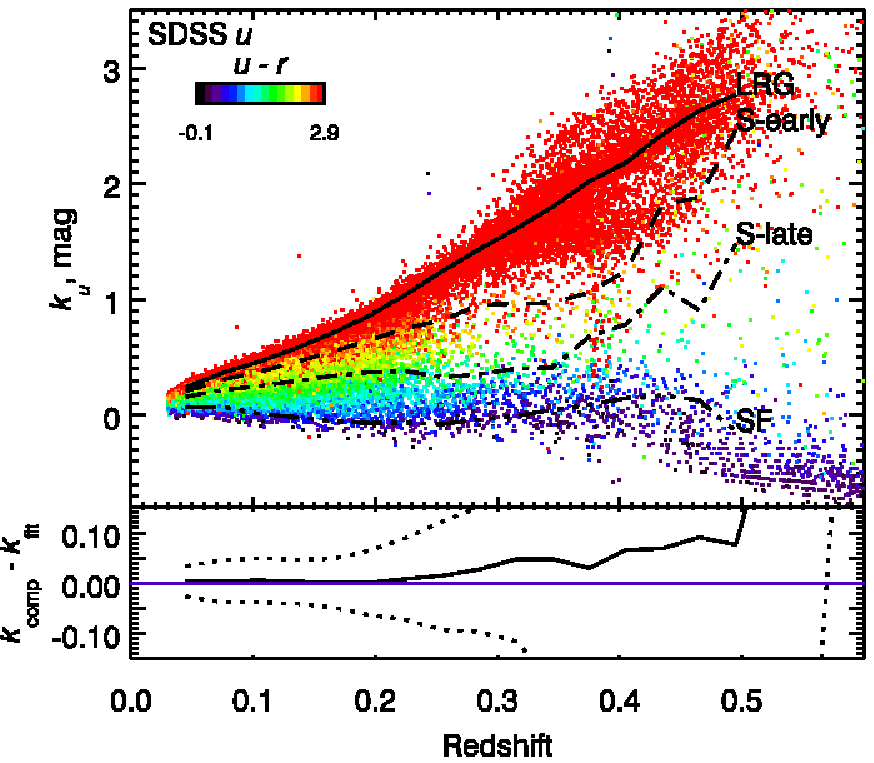}
\includegraphics[width=0.33\hsize]{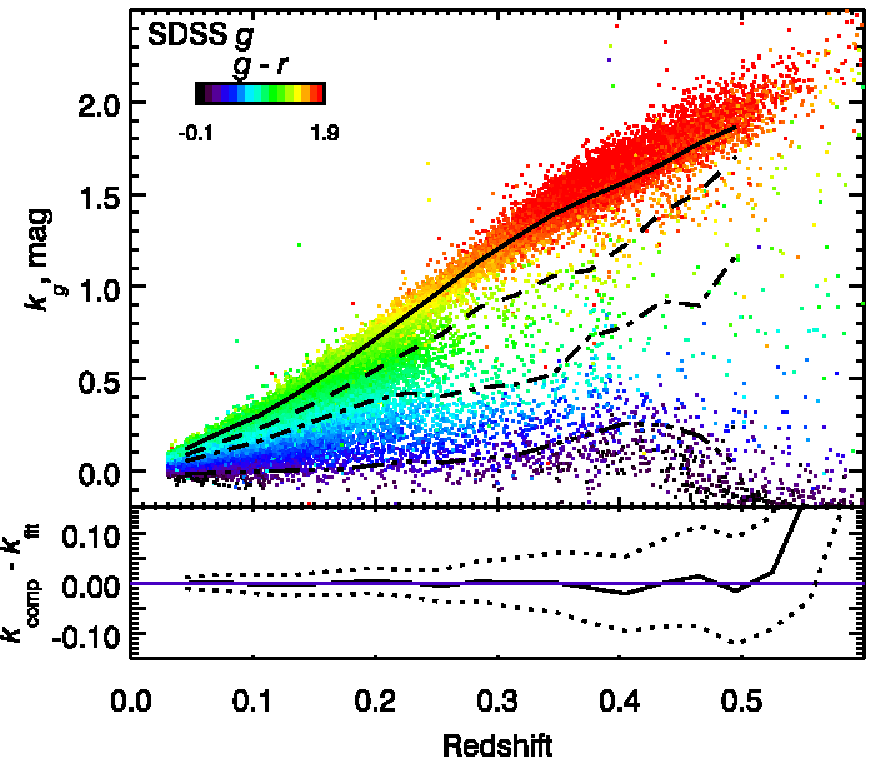}
\includegraphics[width=0.33\hsize]{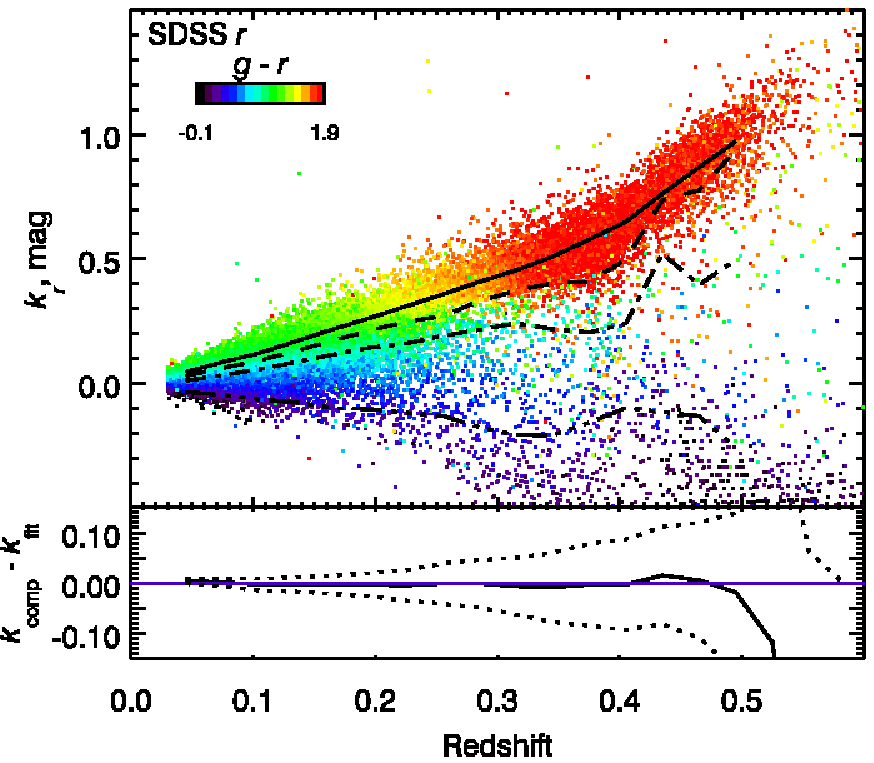}\\
\includegraphics[width=0.33\hsize]{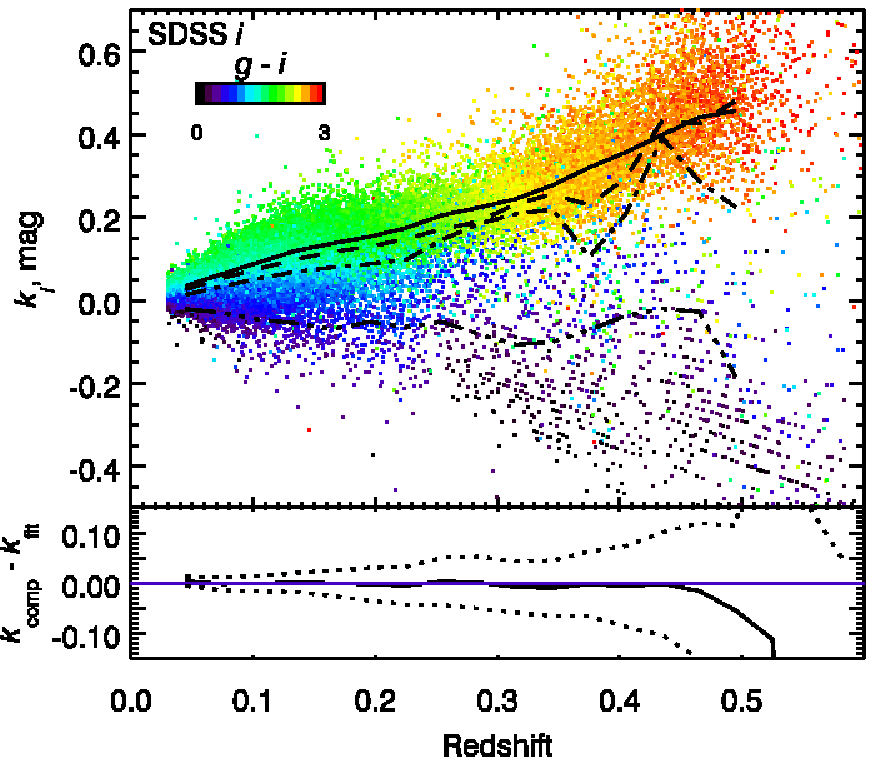}
\includegraphics[width=0.33\hsize]{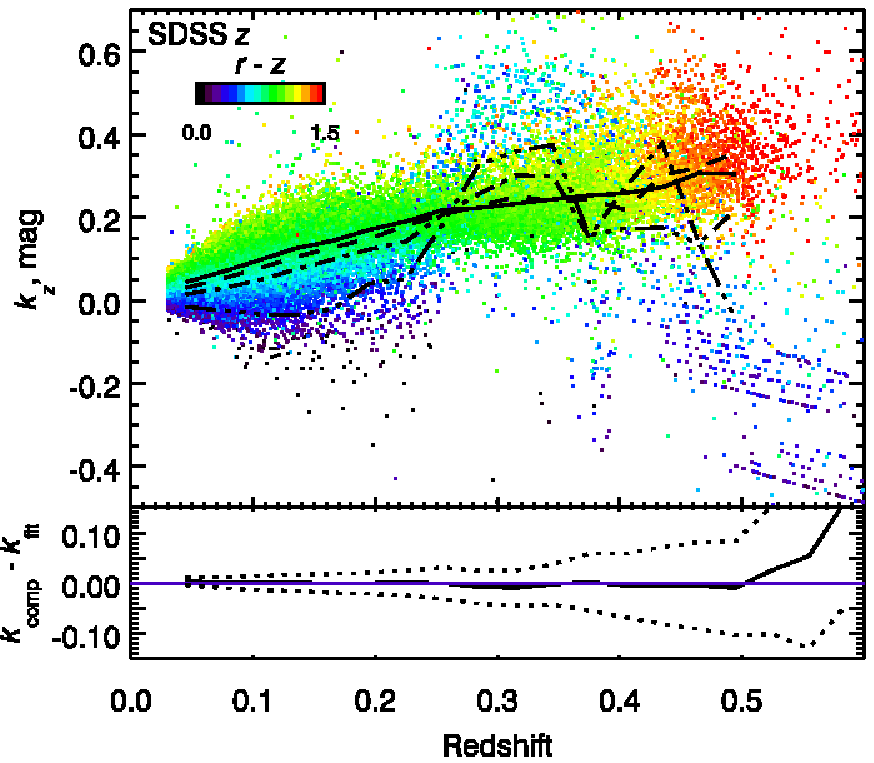}
\includegraphics[width=0.33\hsize]{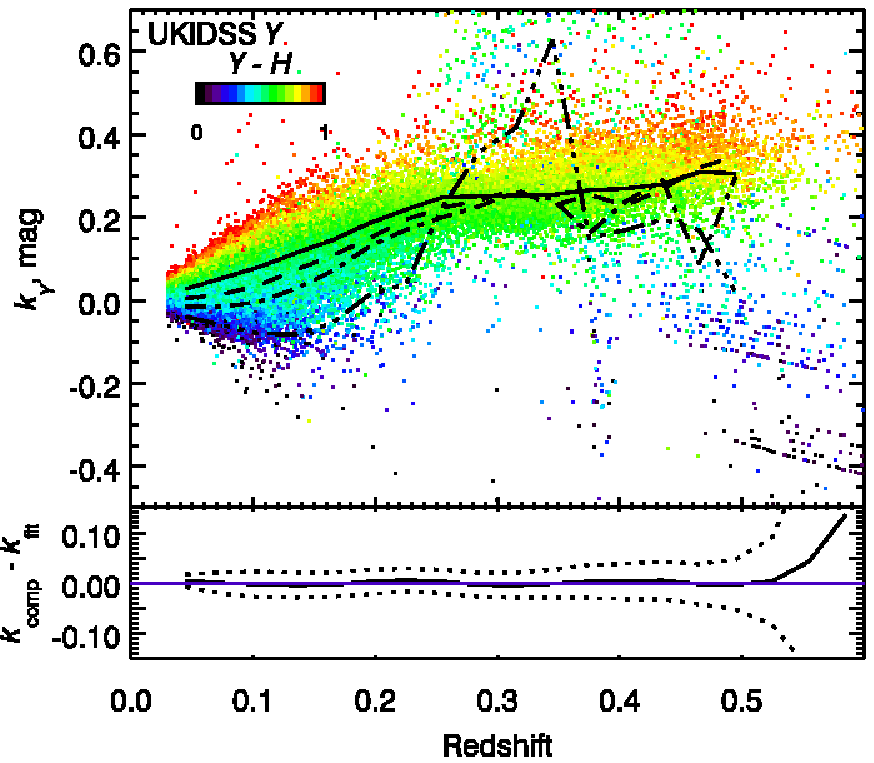}\\
\includegraphics[width=0.33\hsize]{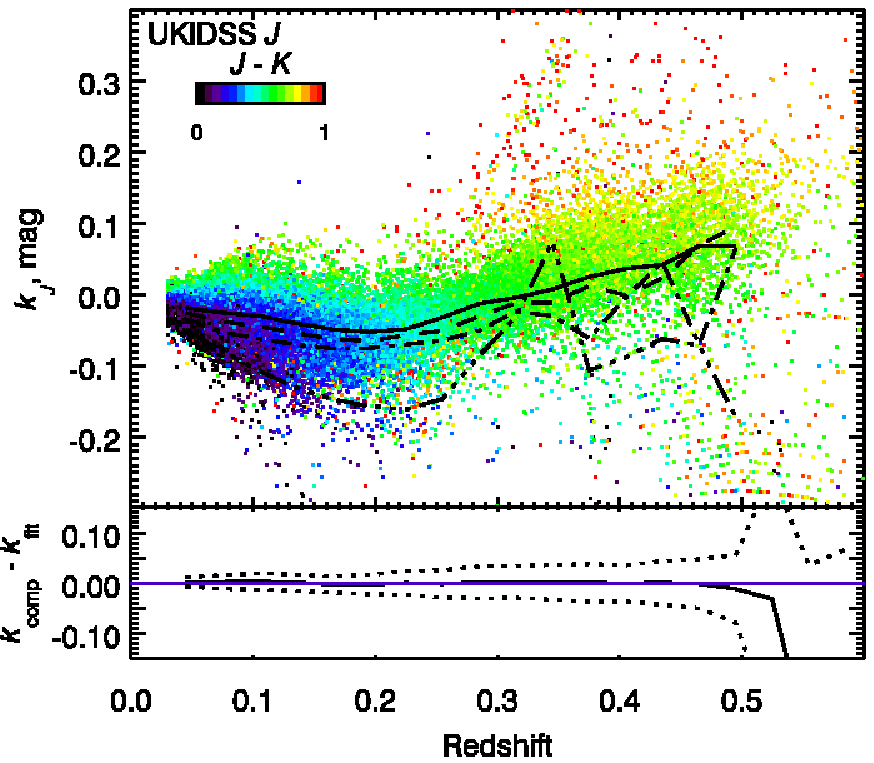}
\includegraphics[width=0.33\hsize]{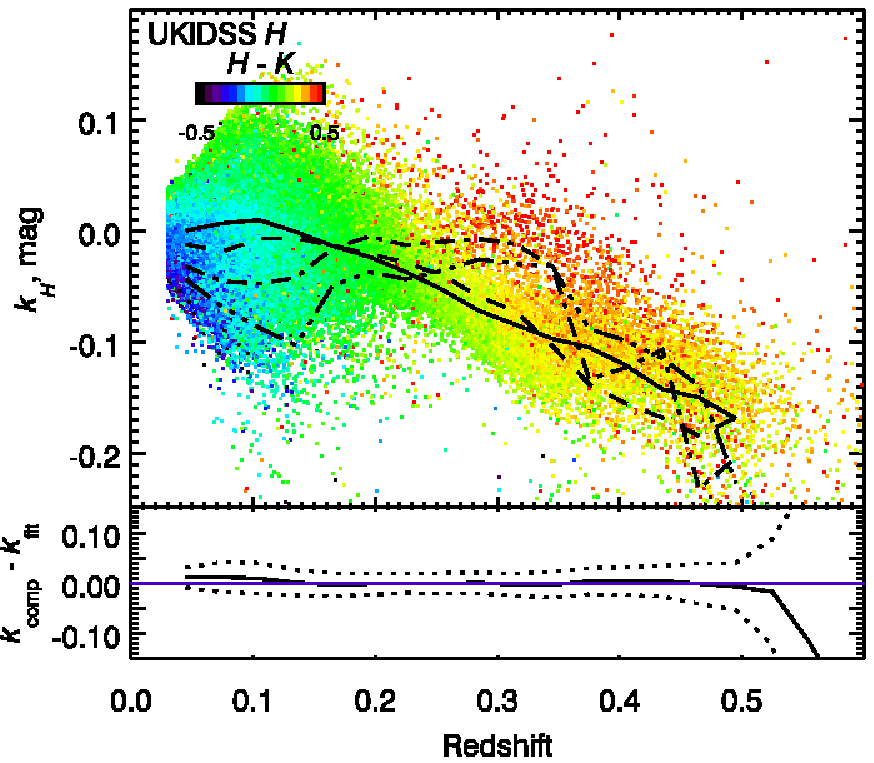}
\includegraphics[width=0.33\hsize]{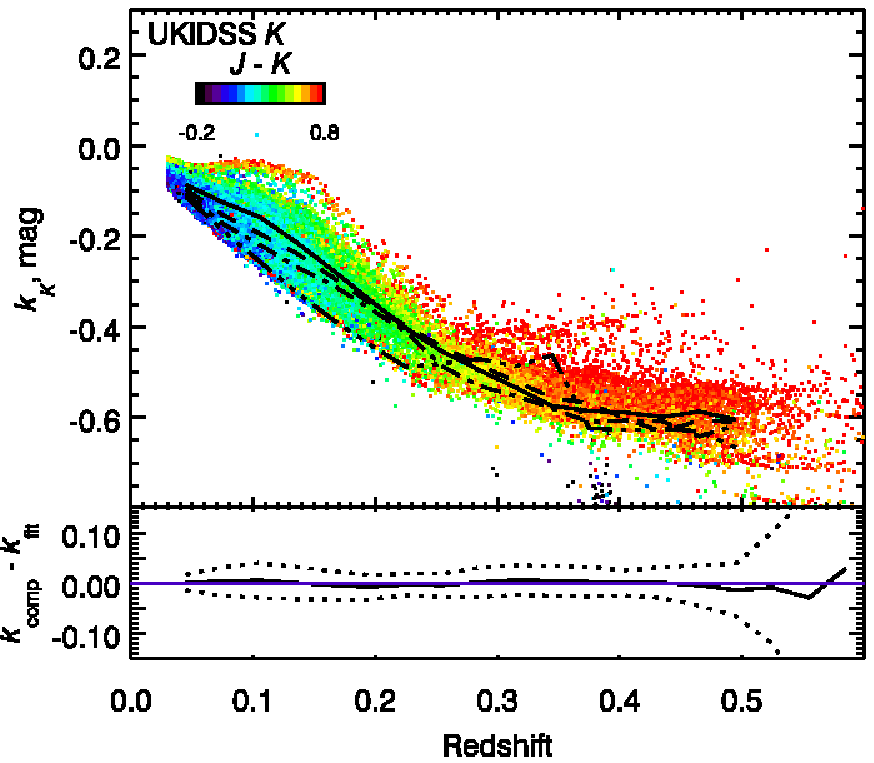}\\
\caption{$K$-corrections in 9 photometric SDSS-UKIDSS bands computed using {\sc pegase.2} SSP 
matching and the residuals of the analytical fitting. Upper panel for
each filter presents computed $K$-corrections vs redshift with a
colour-coded observed colour used to perform the fitting. Sequences of
galaxies with constant restframe $g - r$ colours roughly corresponding
to different morphological types are overplotted (see the text for
details). Lower panels display the residual between the analytical
approximation and measured values (bold black line) and r.m.s. of the
fitting residuals (dotted lines).\label{figKcorrP2fit}}
\end{figure*}

\begin{figure*}
\includegraphics[width=0.33\hsize]{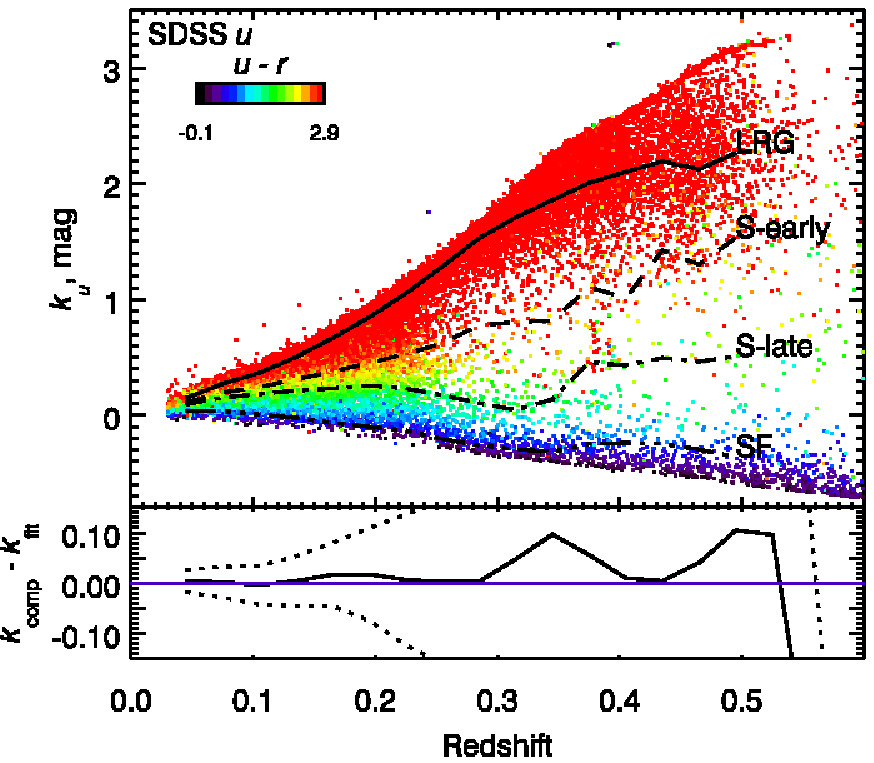}
\includegraphics[width=0.33\hsize]{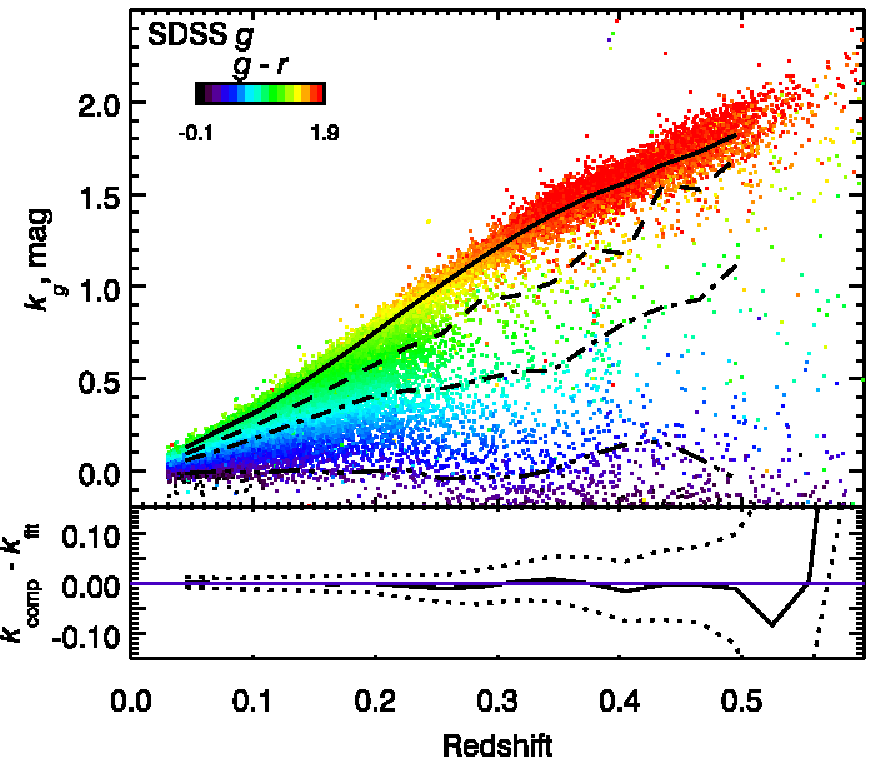}
\includegraphics[width=0.33\hsize]{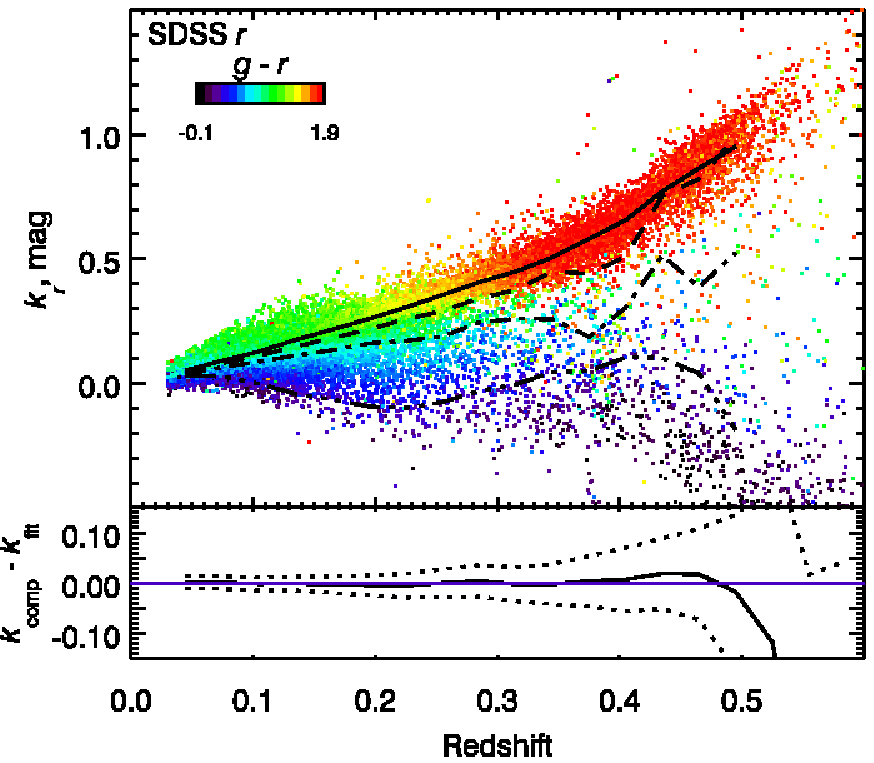}\\
\includegraphics[width=0.33\hsize]{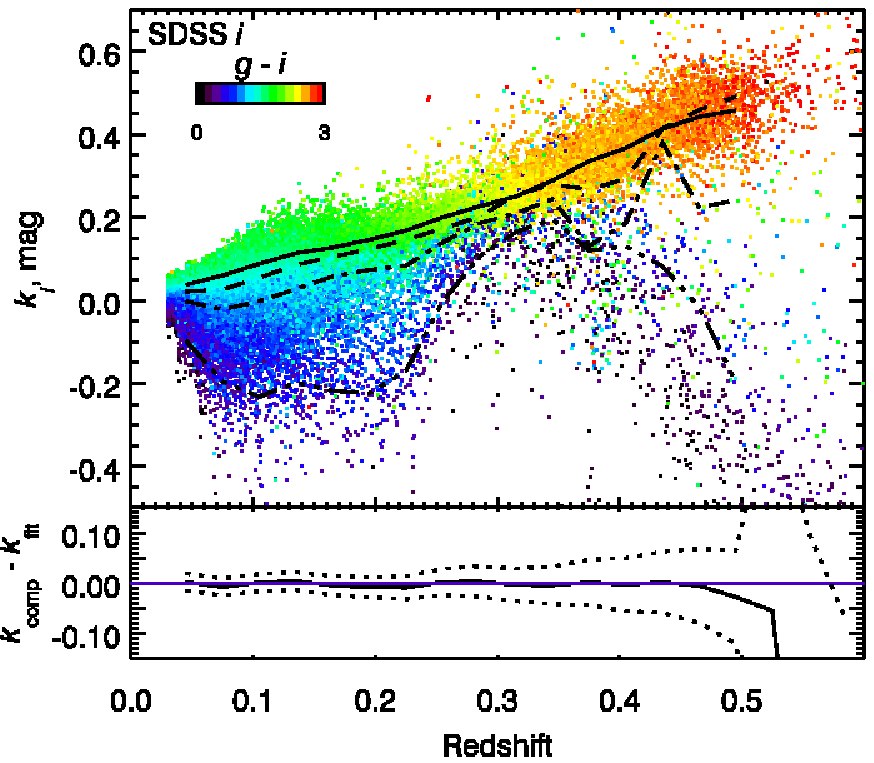}
\includegraphics[width=0.33\hsize]{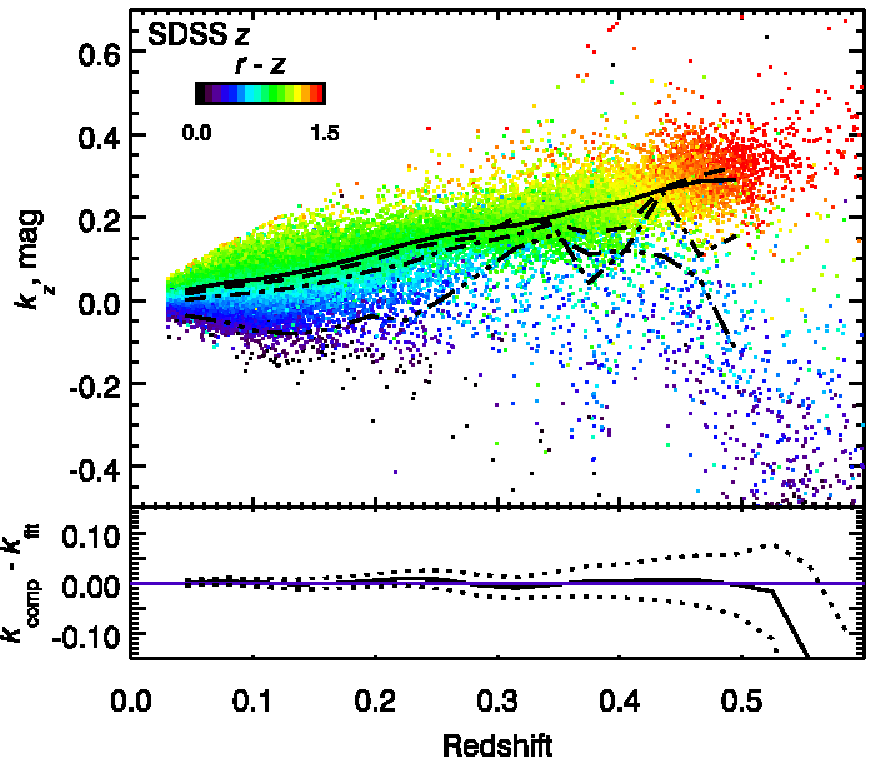}
\includegraphics[width=0.33\hsize]{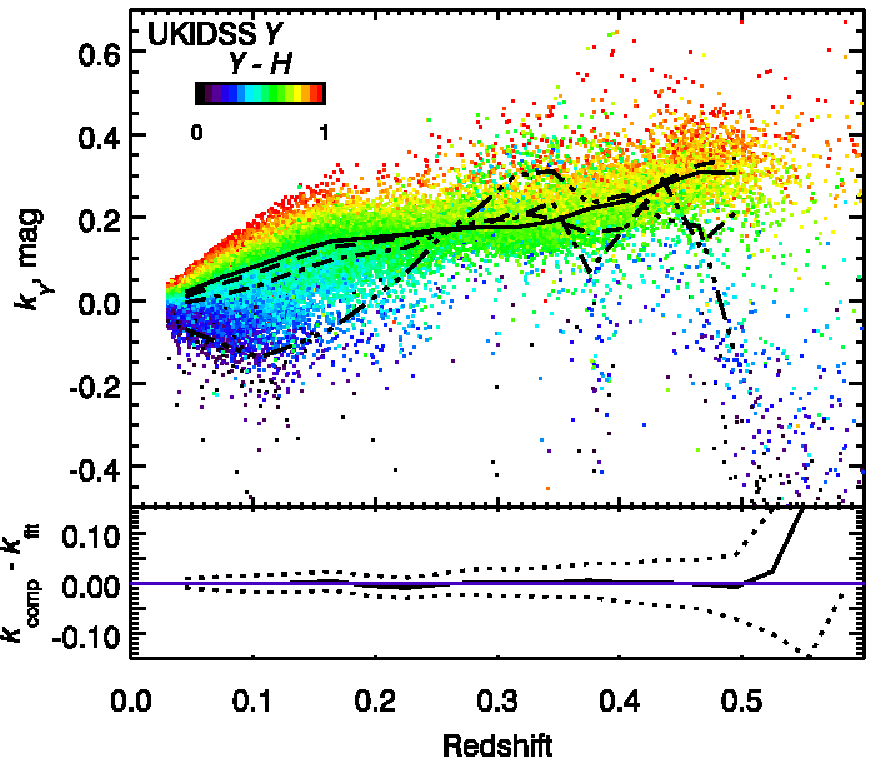}\\
\includegraphics[width=0.33\hsize]{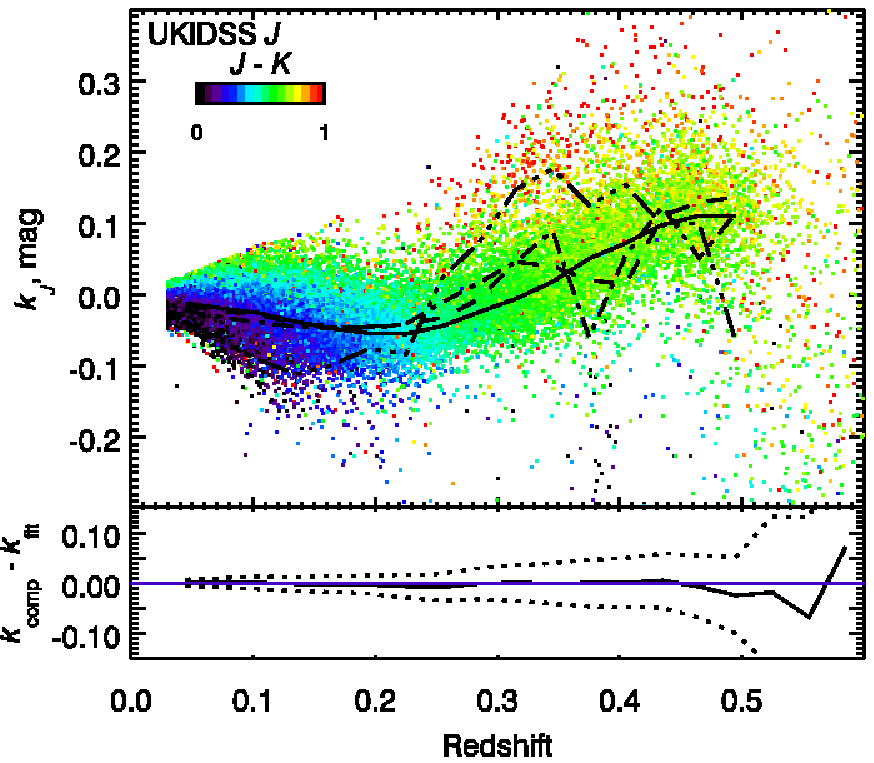}
\includegraphics[width=0.33\hsize]{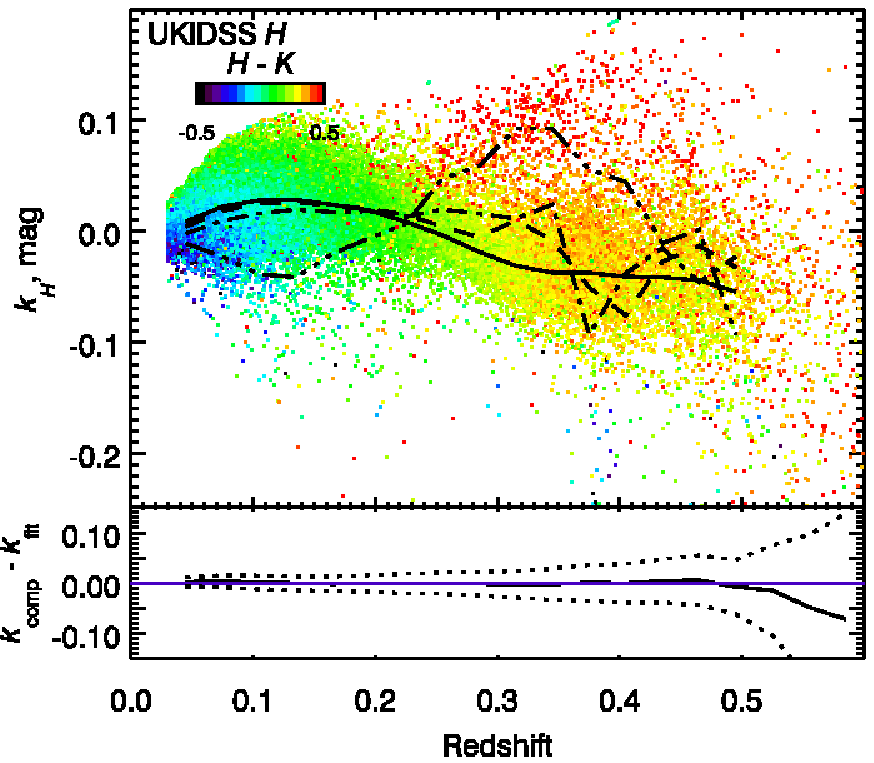}
\includegraphics[width=0.33\hsize]{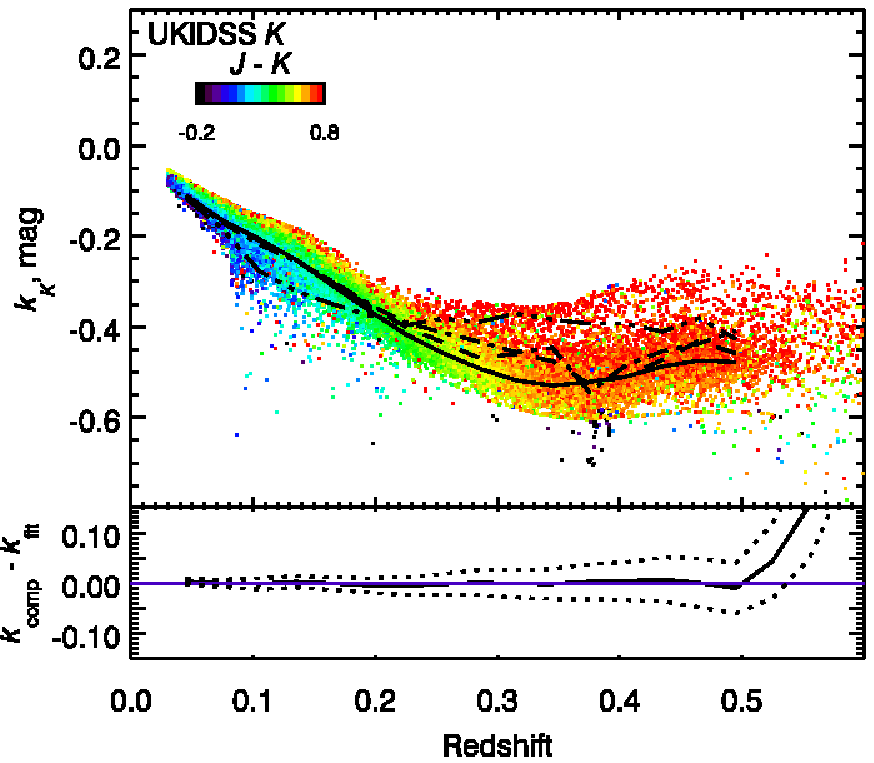}\\
\caption{The same as Fig~\ref{figKcorrP2fit}, but the $K$-corrections are
 computed using the {\sc
kcorrect} code.\label{figKcorrBR07fit}}
\end{figure*}

In Fig~\ref{figKcorrP2fit}--\ref{figKcorrBR07fit}, we display the
analytical approximations of $K$-corrections in 9 bands for the values
computed using {\sc pegase.2}-based matching and the {\sc kcorrect}
code. Upper panels demonstrate computed values as a function of
redshift and an observed colour, colour-coded according to the scale
bars presented in the plots. The lower panels display the mean
residuals and their r.m.s.  as a function of redshift. Since we use a
fixed grid of {\sc pegase.2} templates and do not perform
interpolation on the age and metallicity axes, the individual template
age sequences become visible at high redshifts range in some filters
(e.g. $Y$). However, the computational errors due to this
discretisation do not exceed 0.03~mag. Black lines in the plots denote
the behaviour of $K$-corrections for galaxies having fixed restframe
(i.e. $K$-corrected) $g - r$ colours. Solid lines are for $0.73 < g -
r < 0.81$~mag, dashed for $0.58 < g - r < 0.70$~mag, dot-dashed for
$0.4 < g - r < 0.6$~mag, and triple-dot-dashed for $g - r < 0.15$~mag,
roughly corresponding to luminous red galaxies (LRG), early-type
spirals (S-early), late-type spirals (S-late), and actively
star-forming galaxies (SF). These lines are constructed by connecting
median values within 0.03-wide redshift bins in the colour bins given
above. We notice that the $u$-band $K$-corrections computed with the
{\sc kcorrect} code at higher redshifts turn to be higher than
expected for LRGs defined by $g - r$ restframe colours, which is
probably indicative of a template mismatch.

The 4 morphological types are well separated in the $u$, $g$, and $r$ bands;
only very blue objects exhibit a distinguished behaviour in $i$, while in
NIR bands the $K$-corrections become nearly independent of a restframe
colour of a galaxy (or its morphological type).

Some tables containing the coefficients of polynomial approximations are
provided in Appendix~A. Additional tables for $K$-correction 
approximations using different observed colours are provided on
the ``K-corrections calculator'' web-site described in Appendix~B.

Classical Johnson-Cousins photometric system is still widely used for
extragalactic research, as well as NIR bands of the 2MASS survey
\citep{Skrutskie+06}, therefore computation and analytical approximation of
$K$-corrections in these bands are of a great practical
importance. However, we do not have photometric measurements for
galaxies in our sample obtained in these bands. We used the
photometric transformations defined in
\citet{JGA06} for stars and those provided on the web-pages of SDSS in order to
convert available SDSS $ugriz$ magnitudes into Johnson-Cousins
$UBVR_cI_c$, and UKIDSS $YJHK$ to 2MASS $JHK_s$ transformations
presented in
\citet{HWLH06}. Then, we used these magnitudes to fit $K$-corrections in
the $UBVR_cI_c$ and 2MASS $JHK_s$ bands as functions of ``computed''
observed colours defined in the same photometric system. The coefficients of
these polynomial approximations are available at the ``K-corrections
calculator'' web-site. We note that the approximations for the Johnson--Cousins and
2MASS filters are provided as functions of colours in the Vega system, 
whereas for SDSS and UKIRT WFCAM bands the colours are expressed in AB magnitudes.

\subsection{Validation using SDSS DR7 spectra}

We fetched optical SDSS spectra in the wavelength range between 3800 and
9200\AA\ for all galaxies from our sample, therefore we were able to perform
independent direct verification of the analytical approximations of
$K$-corrections presented above. The available spectral coverage allows us
to compute fluxes directly from the spectra using both real and redshifted
filter transmission curves, i.e. to obtain real restframe fluxes, for $g$,
$r$, and $i$ filters. The highest redshifts, where reliable computation is
still possible for the $r$ and $i$ bands are $\sim 0.28$ and $\sim 0.08$
respectively due to the upper wavelength limit of the SDSS spectral
coverage. We use a similar technique to \citet{RBH09} for the computation of
spectral-based $K$-corrections, with two main differences: (1) we redshift
the filter transmission curves instead of blueshifting the spectra in order
to prevent interpolation of fluxes and smoothing defects in the spectra; (2)
in case of missing data within the wavelength range, we interpolate fluxes
linearly or extrapolate them using constant level in $F_{\lambda}$ if the
red tail of the redshifted filter transmission curve goes beyond 9200\AA.
The maximal adopted truncation of the filter transmission curve was at a
level of 1~per~cent of the maximal transmission.

\begin{figure}
\includegraphics[width=\hsize]{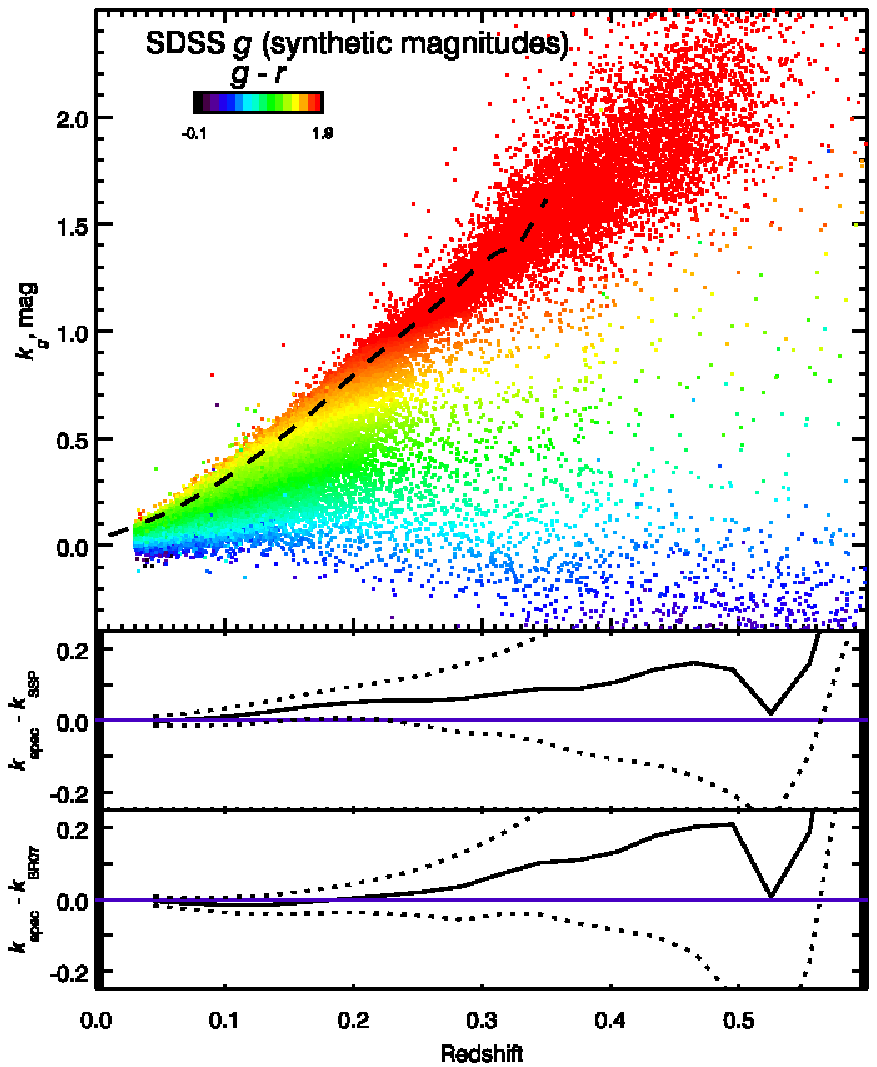}\\
\includegraphics[width=\hsize]{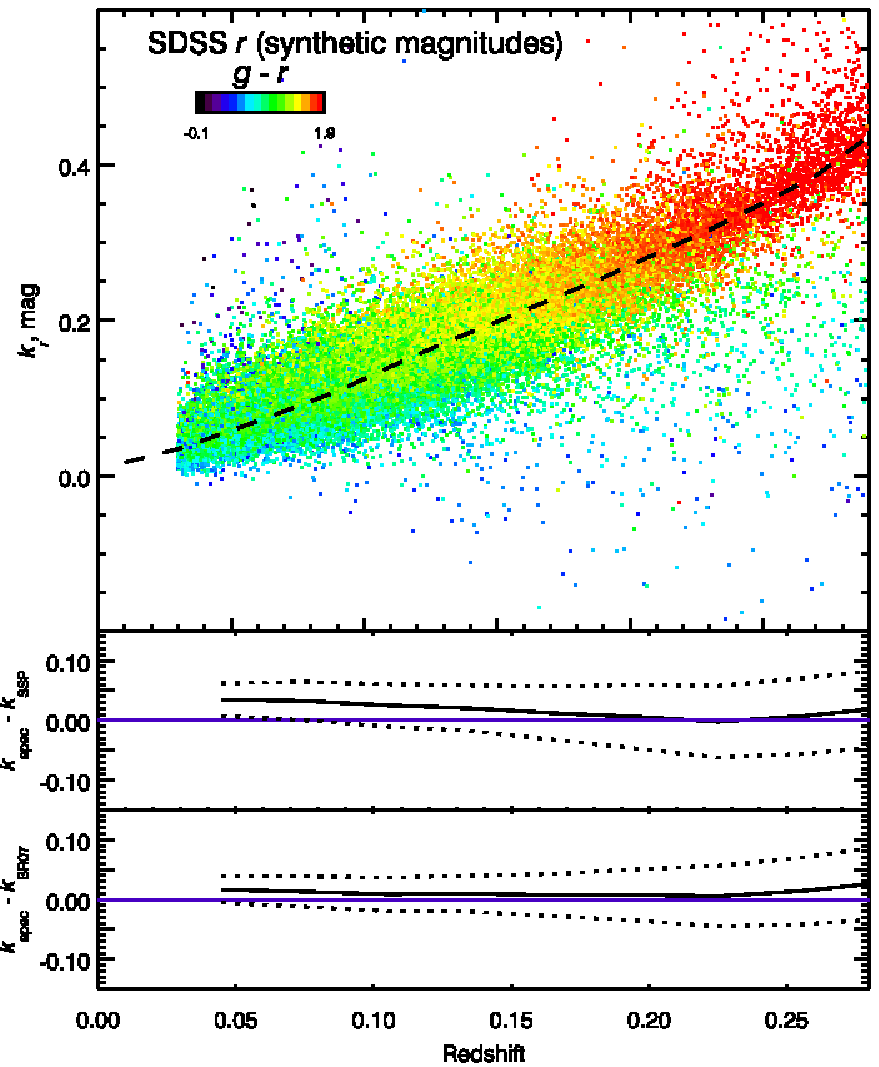}
\caption{Differences between analytically approximated and directly measured spectral-based
$K$-corrections in the $g$ (top) and $r$ (bottom) bands. Upper panels
display computed values of $K$-corrections. Dashed lines denote
spectral-based $K$-corrections for early-type galaxies presented in
\citet{RBH09}. Two bottom panels 
for each spectral band display the differences between spectral-based
$K$-corrections and those computed using {\sc pegase.2} SSP matching and
the {\sc kcorrect} code, respectively. \label{figspecKcorr}}
\end{figure}

In Fig~\ref{figspecKcorr}, we present the differences between spectral-based
$K$-corrections in $g$ and $r$ bands and the analytical approximations of
those computed empirically from SDSS \textit{fiberMag} using {\sc pegase.2}
SSP templates and the {\sc kcorrect} code as functions of redshift and
observed $g - r$ colour. In the $r$ band, the agreement between
spectral-based $K$-corrections and analytical fitting derived above is as
good as 0.02~mag. For the low-redshift ($Z < 0.3$) part of the sample, the
situation is similar in the $g$ band, although the systematic difference
reaches 0.06~mag. However, for the higher redshifts ($0.3 < Z < 0.5$), the
discrepancies become as large as 0.15--0.2~mag. This inconsistency possibly
originates from underestimated $g$ synthetic magnitudes due to a very low
signal in the blue part of the spectra for objects having $Z > 0.3$.
Formally computed photometric errors of synthetic $g$ magnitudes in this
redshift range reach 0.2~mag.

\section{Discussion and conclusions}

\subsection{Comparison with literature}

We compare $K$-corrections computed in this study with the results described
in literature. \citet{FSI95} presented $K$-corrections of the SDSS
photometric system in their fig.~20. Since no data are provided in the
numerical form, we are able to compare the results only qualitatively. The
behaviour of the values as functions of redshifts agree well at $Z<0.5$,
different morphological types of galaxies in \citet{FSI95} are related to
the reconstructed restframe colours. The elliptical
galaxy template of \citet{FSI95} behaves similarly to the sequences
denoted as ``LRG'' in Fig.~\ref{figKcorrP2fit}--\ref{figKcorrBR07fit}
corresponding to luminous red galaxies. We notice that the filter
transmission curves used in \citet{FSI95} are somewhat different from those
obtained from the telescope and presented at the SDSS DR7 web pages,
especially for the $g$ and $r$ bands, explaining why the exact match between
the two approaches cannot be achieved.

\citet{Mannucci+01} present in their fig.~7 NIR $K$-corrections computed from
the templates spectra of galaxies having different morphological types
computed with the {\sc pegase} code. Although their $JHK$ bands  are
slightly different from those of the UKIDSS, the behaviour of
$K$-corrections as functions of redshift is very similar. Authors mention
that $K$-corrections in $H$ and $K$ are virtually insensitive to the
galaxy morphological types, which we observe as their very weak dependence
on galaxy colours.

We compare our results with the spectral-based $K$-corrections presented in
\citet{RBH09}. The authors deal only with red sequence galaxies, therefore
we are able to test only a small although very important part of the
parameter space. The values of $K$-corrections provided in tables~1--2 of
\citet{RBH09} are overplotted in Fig~\ref{figspecKcorr}. The agreement is
remarkably good, which is, however, expected because our spectral-based
$K$-corrections computed in the same manner agree well with the best-fitting
solutions.

\subsection{Recovered restframe magnitudes and LRGs}

\begin{figure}
\includegraphics[width=\hsize]{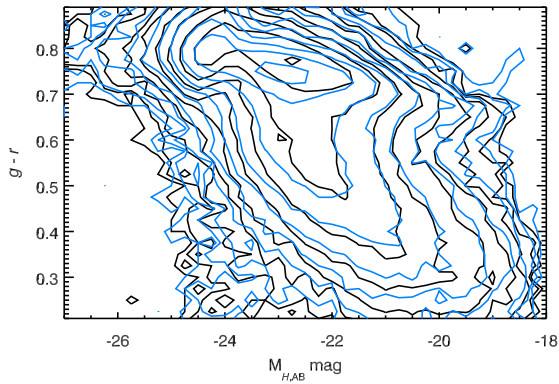}
\caption{Colour-magnitude diagram presenting data for 74,254 galaxies 
using the SDSS DR7 and UKIDSS DR5 Petrosian magnitudes in the $g$, $r$, and 
$H$ bands respectively $K$-corrected using the analytical approximations
presented in this study. Black and blue contours are for the {\sc
pegase.2}-based and {\sc kcorrect} derived values respectively. Levels of
the two-dimensional density plot correspond to the powers of two from 
2 (outermost) to 1024 (innermost) galaxies per bin of 0.25~mag$\times$0.025~mag.
\label{RSplot}}
\end{figure}

It is well known that the restframe colours correlate with the galaxy
morphology, and there is a pronounced bimodality in the colour distribution
of galaxies \citep[see e.g.][]{Strateva+01,BGB04,BBN04}. The so-called ``red
sequence'' includes elliptical and lenticular galaxies with some fraction of
early-type spirals demonstrating a lack of ongoing star formation,
whereas the ``blue cloud'' contains actively star-forming objects
represented mostly by late-type spiral and irregular galaxies. Some
morphologically classified early-type galaxies indeed sit below the red
sequence in the ``green valley'' or even ``blue cloud'' regions. In
Fig~\ref{RSplot}, we provide the colour-magnitude plot of 74,254 galaxies
from SDSS DR7 and UKIDSS, where their total Petrosian magnitudes were
$K$-corrected using the presented analytical approximations.

Some of the objects residing below the red sequence referred as E+A galaxies
\citep{DG87,CS87,DSP99} often exhibit weak if any emission lines in their
spectra ruling out major star formation. However, Balmer absorptions
are remarkably deep, which is a good evidence for the presence of
young stars that was confirmed by the recent study where authors used
full-spectral fitting \citep{CDRB09}. Despite the early-type
morphologies of E+A, these galaxies are indistinguishable from spirals if
one uses only integrated colours and therefore, colour transformations and
$K$-corrections behave very similarly for these two morphologically different
classes of galaxies. Hence, the restframe colours are not tightly connected
to the galaxy morphology, but to the internal properties such as ongoing
star formation and, therefore, they are specific of stellar populations. Another example
would be dwarf elliiptical (dE) and compact elliptical (cE) galaxies. The
former ones usually have intermediate-age stellar populations with
metallicities lower than those of LRGs \citep[see
e.g.][]{Chilingarian+08,Chilingarian09}, therefore their global optical
colours are bluer, and their $K$-corrections in the corresponding bands are
lower than those of more massive early-type galaxies, which results in quite
a strong tilt of the red sequence at $M_{H,AB}
> -20$~mag (see Fig~\ref{RSplot}) reproducing its behaviour for nearby dE galaxies
\citep[][]{JL09}. On the other hand, cE galaxies with luminosities similar
to dEs, originate from the tidal
stripping of more massive early-type progenitors and, therefore, exhibit
old metal-rich stellar populations \citep{Chilingarian+09}. For this reason,
cEs have significantly redder colours compared to dEs and, therefore, have
behaviour of $K$-corrections very similar to LRGs.

There is a weak dependence of the colour of red sequence galaxies on the
galaxy luminosity reflecting, in particular, a mass-metallicity relation of
early-type galaxies. Massive (and luminous) red sequence galaxies compared
to lower-mass systems contain more metal-rich stellar populations having
intrinsically redder colours than the metal-poor ones.

The special case of LRGs is very important for understanding galaxy
formation and evolution, therefore we provide specific approximations for
these objects. Since the intrinsic spread of LRGs restframe colours is very
low, about 0.04~mag, their $K$-corrections can be well approximated as a
function of a single parameter, the redshift. We selected a sample of LRGs
based on their restframe colours, and fitted the $K$-corrections as
polynomial functions of their redshifts in all 9 photometric bands. The
coefficients of best-fitting 5th order polynomials with no constant term are
presented in Table~\ref{tabLRG1} and Table~\ref{tabLRG2} for the {\sc
pegase.2} SSP-based and {\sc kcorrect}-computed values correspondingly.
Every row corresponds to a given photometric band and contains coefficients
from the 1st to the 5th power of redshift.

\begin{table}
\caption{Coefficients of the best-fitting polynomials for the
redshift-dependence of $K$-corrections for luminous red galaxies. Initial
$K$-correction values were computed from the {\sc pegase.2} SSPs.
\label{tabLRG1}}
\begin{tabular}{lccccc}
 & $Z^1$ & $Z^2$& $Z^3$& $Z^4$& $Z^5$ \\
\hline
$K_u$ &      5.93938&     -30.5247&      179.473&     -380.488&      282.011\\
$K_g$ &      2.61617&     -4.44391&      93.0132&     -284.582&      252.245\\
$K_r$ &     0.312233&      14.3325&     -68.2493&      136.254&     -87.3360\\
$K_i$ &     0.234538&      14.3162&     -97.2754&      246.775&     -207.028\\
$K_z$ &     0.897075&      3.60112&     -34.7890&      93.0266&     -79.5246\\
$K_Y$ &     0.402992&      5.30858&     -18.3172&      17.6760&     -0.31400\\
$K_J$ &    -0.076704&     -4.48411&      27.1585&     -45.6481&      22.6928\\
$K_H$ &     0.382926&     -1.81590&     -13.1657&      57.5486&     -59.0677\\
$K_K$ &     -1.75997&      5.48023&     -56.4175&      175.939&     -160.754\\
\hline
\end{tabular}
\end{table}

\begin{table}
\caption{Same as in Table~\ref{tabLRG1} but the {\sc kcorrect} computed
$K$-corrections.
\label{tabLRG2}}
\begin{tabular}{lccccc}
 & $Z^1$ & $Z^2$& $Z^3$& $Z^4$& $Z^5$ \\
\hline
$K_u$ &      4.20000&     -24.5015&      229.149&     -574.272&      434.901\\
$K_g$ &      2.17470&      10.3810&      1.49141&     -76.6656&      88.6641\\
$K_r$ &     0.710579&      10.1949&     -57.0378&      133.141&     -99.9271\\
$K_i$ &     0.702681&      4.27115&     -37.2060&      112.054&     -105.976\\
$K_z$ &     0.643953&     -1.88400&      13.6952&     -35.0960&      29.9249\\
$K_Y$ &    -0.245996&      21.8772&     -137.019&      322.051&     -257.136\\
$K_J$ &     0.106358&     -5.06024&      18.4707&     -3.73196&     -23.9595\\
$K_H$ &     0.268479&      3.03488&     -35.8994&      98.6524&     -83.9401\\
$K_K$ &     -2.80894&      15.6923&     -96.8401&      256.235&     -220.691\\
\hline
\end{tabular}
\end{table}

\begin{figure}
\includegraphics[width=\hsize]{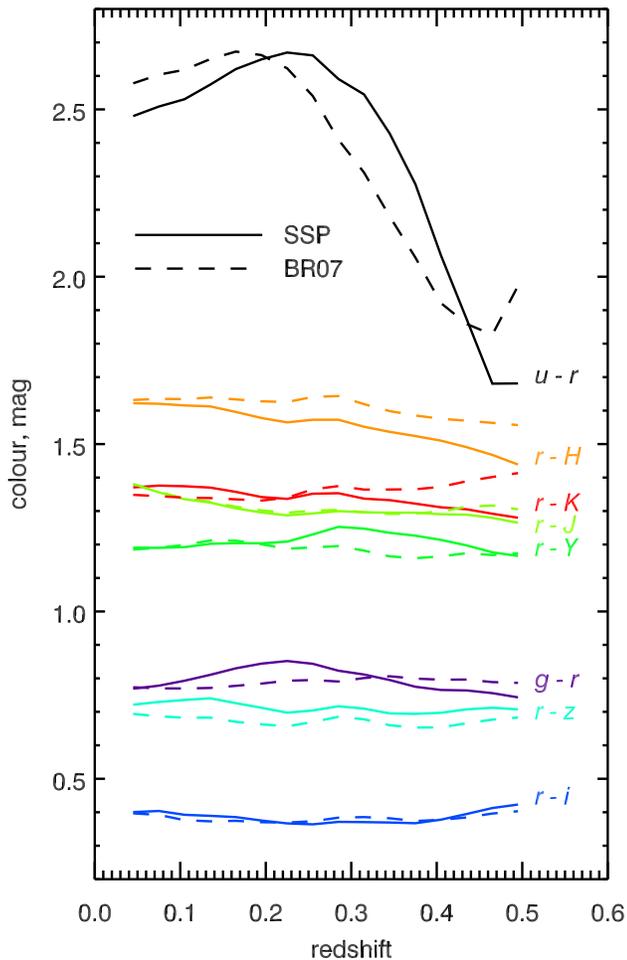}
\caption{Dependence of recovered restframe colours of LRGs on redshift.
Solid and dashed lines are for the {\sc pegase.2} SSP-based and {\sc
kcorrect} computed $K$-corrections respectively.\label{figLRG}}
\end{figure}

Rest-frame colours of LRGs provide a natural test for the quality of
$K$-corrections. The vast majority of stars in these objects is believed to form
on short timescales in the early Universe and then evolve passively. The
redshift $z=0.5$ corresponds to the lookback time about 5~Gyr. Thus, if one
assumes LRGs to be as old as 12~Gyr in the local Universe and contain no
younger stars, then at $z=0.5$ their restframe optical colours would be
slightly bluer with the differences $\Delta(u - r) \approx 0.25$~mag,
$\Delta(g - r) \approx 0.08$~mag, and redder colours different by less than
0.03~mag (values estimated from the colour evolution of the solar
metallicity {\sc pegase.2} SSP). In Fig~\ref{figLRG}, we present the
behaviour of recovered restframe colours of LRGs as functions of redshift.
All presented colours have SDSS $r$ as one of the bands, because the
best consistency of both $K$-correction computation algorithms 
is reached in this band.

The bluest $u - r$ colour exhibits significant changes over redshift
exceeding by a factor of 3 the expectations from the passive evolution of a
SSP. This, however, can be explained by a ``tail'' of the star formation
history since even small mass fractions of intermediate mass stars strongly
affect the $u$ band photometry. On the other hand, we cannot exclude the
template mismatch between the model and real galaxies to be partly
responsible for this effect.

Surprisingly, the $g - r$ colour does not evolve at all if we use the {\sc
kcorrect} values of $g$-band $K$-corrections, but evolves slightly stronger
than expected from the SSP evolution when using the {\sc pegase.2} based
values, being somewhat consistent with the $u - r$ colour behaviour.

While $r - i$ and $r - z$ colours behave similarly in both approaches
exhibiting, as expected, virtually no evolution, in NIR bands the
situation is different. Both techniques are consistent in the $J$
band, although demonstrating rather unexpected evolution by about
0.08~mag, whereas in $H$ and $K$ they are significantly different (as
also can be seen from the corresponding panels in
Fig.~\ref{figcompKcorr}). The {\sc pegase.2} SSP-based $K$-band
$K$-corrections at $Z>0.3$ look slightly underestimated, conversely to
the {\sc kcorrect}-computed ones, which are overestimated. In the $H$
band, both methods seem to underestimate $K$-corrections, but
SSP-based approach is stronger affected.

The decisive answer to the questions about the $K$-correction computation in
NIR bands will be given only when the next generation NIR stellar population
models based on empirical stellar libraries become available. The $H$ and
$K$ band solutions presented in this paper have to be used with caution
keeping in mind that some systematic errors may be introduced to the final
results.

\subsection{Summary}

We present polynomial approximations of $K$-corrections in commonly used
optical (SDSS $u g r i z$ and Johnson-Cousins $UBVR_cI_c$) and
near-infrared (UKIRT WFCAM $Y J H K$ and 2MASS $J H K_s$) broad-band
filters as functions of redshift and observed colours for galaxies at
redshifts $Z<0.5$. The traditional $K$-correction computation techniques
based on the SED fitting require multi-colour photometry, which is not
always available. Our approach allows one to compute restframe galaxy
magnitudes of the same quality using a minimal set of observables including
only two photometric points and a redshift. For luminous red galaxies, we
provide the fitting solutions of $K$-corrections as 1-dimensional polynomial
functions of a redshift.

\section*{Acknowledgments}

In this study, we used the UKIDSS DR5 survey catalogues available through the
WFCAM science archive and SDSS DR7 data. Funding for the SDSS and SDSS-II
has been provided by the Alfred P. Sloan Foundation, the Participating
Institutions, the National Science Foundation, the U.S. Department of
Energy, the National Aeronautics and Space Administration, the Japanese
Monbukagakusho, the Max Planck Society, and the Higher Education Funding
Council for England. The SDSS Web Site is http://www.sdss.org/. We
acknowledge the usage of the {\sc topcat} software by M.Taylor. IC and IZ
acknowledge the support from the RFBR grants 07-02-0029 and 09-02-00032.
Special thanks to A.~Sergeev for the ``K-corrections calculator'' web-site design.
This research is supported by the VO Paris Data Centre. We thank our
anonymous referee for useful suggestions.

\bibliographystyle{mn2e}
\bibliography{K_corr}

\appendix

\section{2-dimensional approximations of $K$-corrections}

Here we provide tables containing coefficients of the analytical
approximations of $K$-corrections. Columns in every table contain colour
terms of the fitting solution, from constant to the 3rd order respectively.
Rows are for the redshift dependence. The maximal degree of the polynomial
surface is set to 5 or 7 (for the $H$ band), therefore the coefficients for
the higher degrees, in both colour and redshift, are set to zero. Tables
\ref{tabu_r}--\ref{tabKJ_K} and \ref{br07tabu_r}--\ref{br07tabKJ_K} present the 
coefficients for approximations of {\sc pegase.2}-based and {\sc
kcorrect}-computed $K$-corrections in the SDSS $ugriz$ and UKIDSS $YJHK$ bands. Tables
\ref{tabU_R_c}--\ref{tabI_c_V_I_c} contain the coefficients for
approximations of {\sc pegase.2}-based $K$-corrections in the
Johnson--Cousins $UBVR_cI_c$ bands. The coefficients for the 2MASS $JHK_s$
bands as well as for the other colour combinations of SDSS and UKIDSS bands
are provided online through the ``K-corrections calculator'' service.

\begin{table}
\caption{
The coefficients $a_{x,y}$ of the two-dimensional polynomial approximation
of $K$-corrections $K_{u}(Z, u - r)$ in the SDSS $u$ band (see Equation~1).
In order to derive a value for $K_{u}$, one needs to sum the polynomial terms
of a form $Z^x$ and $(u - r)^y$ from the table heading and its first column
multiplied by the coefficients in corresponding table cells.\label{tabu_r}}

\begin{tabular}{l|cccc}
 & $(u - r)^0$ & $(u - r)^1$ & $(u - r)^2$ & $(u - r)^3$ \\
\hline
$Z^{0}$ &  0  &  0  &  0  &  0 \\
$Z^{1}$ &  1.63349  &  2.24658  &  0.141845  &  -0.13441 \\
$Z^{2}$ &  -71.84  &  20.4939  &  -3.82771  &  0.789867 \\
$Z^{3}$ &  257.509  &  -42.3042  &  -4.05721  &  0 \\
$Z^{4}$ &  -308.573  &  63.0036  &  0  &  0 \\
$Z^{5}$ &  42.8572  &  0  &  0  &  0 \\
\hline
\end{tabular}
\end{table}

\begin{table}
\caption{Same as in Table~\ref{tabu_r} but for the SDSS $g$ band as a function of redshift and the $g - r$ colour.\label{tabg_r}}
\begin{tabular}{l|cccc}
 & $(g - r)^0$ & $(g - r)^1$ & $(g - r)^2$ & $(g - r)^3$ \\
\hline
$Z^{0}$ &  0  &  0  &  0  &  0 \\
$Z^{1}$ &  -0.900332  &  3.97338  &  0.774394  &  -1.09389 \\
$Z^{2}$ &  3.65877  &  -8.04213  &  11.0321  &  0.781176 \\
$Z^{3}$ &  -16.7457  &  -31.1241  &  -17.5553  &  0 \\
$Z^{4}$ &  87.3565  &  71.5801  &  0  &  0 \\
$Z^{5}$ &  -123.671  &  0  &  0  &  0 \\
\hline
\end{tabular}
\end{table}

\begin{table}
\caption{Same as in Table~\ref{tabu_r} but for the SDSS $r$ band as a function of redshift and the $g - r$ colour.\label{tabrg_r}}
\begin{tabular}{l|cccc}
 & $(g - r)^0$ & $(g - r)^1$ & $(g - r)^2$ & $(g - r)^3$ \\
\hline
$Z^{0}$ &  0  &  0  &  0  &  0 \\
$Z^{1}$ &  -1.61294  &  3.81378  &  -3.56114  &  2.47133 \\
$Z^{2}$ &  9.13285  &  9.85141  &  -5.1432  &  -7.02213 \\
$Z^{3}$ &  -81.8341  &  -30.3631  &  38.5052  &  0 \\
$Z^{4}$ &  250.732  &  -25.0159  &  0  &  0 \\
$Z^{5}$ &  -215.377  &  0  &  0  &  0 \\
\hline
\end{tabular}
\end{table}

\begin{table}
\caption{Same as in Table~\ref{tabu_r} but for the SDSS $i$ band as a function of redshift and the $g - i$ colour.\label{tabg_i}}
\begin{tabular}{l|cccc}
 & $(g - i)^0$ & $(g - i)^1$ & $(g - i)^2$ & $(g - i)^3$ \\
\hline
$Z^{0}$ &  0  &  0  &  0  &  0 \\
$Z^{1}$ &  -2.41799  &  4.68318  &  -3.70678  &  1.5155 \\
$Z^{2}$ &  11.2598  &  5.14198  &  -2.64767  &  -3.63215 \\
$Z^{3}$ &  -94.7387  &  -14.154  &  27.2864  &  0 \\
$Z^{4}$ &  285.775  &  -51.6662  &  0  &  0 \\
$Z^{5}$ &  -222.641  &  0  &  0  &  0 \\
\hline
\end{tabular}
\end{table}

\begin{table}
\caption{Same as in Table~\ref{tabu_r} but for the SDSS $z$ band as a function of redshift and the $r - z$ colour.\label{tabr_z}}
\begin{tabular}{l|cccc}
 & $(r - z)^0$ & $(r - z)^1$ & $(r - z)^2$ & $(r - z)^3$ \\
\hline
$Z^{0}$ &  0  &  0  &  0  &  0 \\
$Z^{1}$ &  -1.7252  &  3.35566  &  0.469411  &  0.350873 \\
$Z^{2}$ &  14.9772  &  -23.1956  &  -3.32427  &  1.78842 \\
$Z^{3}$ &  -41.1269  &  75.2648  &  -10.2986  &  0 \\
$Z^{4}$ &  11.3667  &  -48.3299  &  0  &  0 \\
$Z^{5}$ &  23.5438  &  0  &  0  &  0 \\
\hline
\end{tabular}
\end{table}

\begin{table}
\caption{Same as in Table~\ref{tabu_r} but for the UKIDSS $Y$ band as a function of redshift and the $Y - H$ colour.\label{tabY_H}}
\begin{tabular}{l|cccc}
 & $(Y - H)^0$ & $(Y - H)^1$ & $(Y - H)^2$ & $(Y - H)^3$ \\
\hline
$Z^{0}$ &  0  &  0  &  0  &  0 \\
$Z^{1}$ &  -2.01575  &  2.70429  &  6.01384  &  -5.17119 \\
$Z^{2}$ &  14.3112  &  -13.2354  &  -4.03961  &  13.2633 \\
$Z^{3}$ &  -46.4835  &  12.7464  &  -26.3263  &  0 \\
$Z^{4}$ &  80.2341  &  27.9842  &  0  &  0 \\
$Z^{5}$ &  -66.1958  &  0  &  0  &  0 \\
\hline
\end{tabular}
\end{table}

\begin{table}
\caption{Same as in Table~\ref{tabu_r} but for the UKIDSS $J$ band as a function of redshift and the $J - K$ colour.\label{tabJJ_K}}
\begin{tabular}{l|cccc}
 & $(J - K)^0$ & $(J - K)^1$ & $(J - K)^2$ & $(J - K)^3$ \\
\hline
$Z^{0}$ &  0  &  0  &  0  &  0 \\
$Z^{1}$ &  -0.765217  &  2.43055  &  -0.427304  &  0.277662 \\
$Z^{2}$ &  1.59864  &  -14.646  &  12.0911  &  -1.2131 \\
$Z^{3}$ &  -4.02136  &  18.077  &  -26.1137  &  0 \\
$Z^{4}$ &  18.5608  &  25.2691  &  0  &  0 \\
$Z^{5}$ &  -40.3567  &  0  &  0  &  0 \\
\hline
\end{tabular}
\end{table}

\begin{table}
\caption{Same as in Table~\ref{tabu_r} but for the UKIDSS $H$ band as a function of redshift and the $H - K$ colour.\label{tabH_K}}
\begin{tabular}{l|cccc}
 & $(H - K)^0$ & $(H - K)^1$ & $(H - K)^2$ & $(H - K)^3$ \\
\hline
$Z^{0}$ &  0  &  0  &  0  &  0 \\
$Z^{1}$ &  -0.642942  &  -1.05192  &  -15.5123  &  -18.1957 \\
$Z^{2}$ &  26.3667  &  80.0291  &  192.688  &  179.956 \\
$Z^{3}$ &  -274.11  &  -564.952  &  -848.543  &  -646.653 \\
$Z^{4}$ &  1081  &  1569.47  &  1741.31  &  761.264 \\
$Z^{5}$ &  -1938.48  &  -1893.45  &  -1488.4  &  0 \\
$Z^{6}$ &  1448.38  &  869.396  &  0  &  0 \\
$Z^{7}$ &  -249.952  &  0  &  0  &  0 \\
\hline
\end{tabular}
\end{table}

\begin{table}
\caption{Same as in Table~\ref{tabu_r} but for the UKIDSS $K$ band as a function of redshift and the $J - K$ colour.\label{tabKJ_K}}
\begin{tabular}{l|cccc}
 & $(J - K)^0$ & $(J - K)^1$ & $(J - K)^2$ & $(J - K)^3$ \\
\hline
$Z^{0}$ &  0  &  0  &  0  &  0 \\
$Z^{1}$ &  -2.80374  &  4.14968  &  1.15579  &  -1.94003 \\
$Z^{2}$ &  13.4077  &  -39.5749  &  11.7  &  4.7809 \\
$Z^{3}$ &  -69.7725  &  94.0769  &  -35.1023  &  0 \\
$Z^{4}$ &  157.649  &  -44.0291  &  0  &  0 \\
$Z^{5}$ &  -132.317  &  0  &  0  &  0 \\
\hline
\end{tabular}
\end{table}

\begin{table}
\caption{Same as in Table~\ref{tabu_r} but using the {\sc kcorrect} code.\label{br07tabu_r}}
\begin{tabular}{l|cccc}
 & $(u - r)^0$ & $(u - r)^1$ & $(u - r)^2$ & $(u - r)^3$ \\
\hline
$Z^{0}$ &  0  &  0  &  0  &  0 \\
$Z^{1}$ &  8.81624  &  -12.0027  &  5.57928  &  -0.825005 \\
$Z^{2}$ &  33.0392  &  39.7152  &  -18.7077  &  4.0901 \\
$Z^{3}$ &  -1223.73  &  236.944  &  -52.6404  &  -3.43017 \\
$Z^{4}$ &  6304.63  &  -659.764  &  162.85  &  -1.62112 \\
$Z^{5}$ &  -15288.5  &  395.301  &  -83.6382  &  0 \\
$Z^{6}$ &  18533.6  &  -67.5999  &  0  &  0 \\
$Z^{7}$ &  -8719.48  &  0  &  0  &  0 \\
\hline
\end{tabular}
\end{table}

\begin{table}
\caption{Same as in Table~\ref{br07tabu_r} but for the SDSS $g$ band as a function of redshift and the $g - r$ colour.\label{br07tabg_r}}
\begin{tabular}{l|cccc}
 & $(g - r)^0$ & $(g - r)^1$ & $(g - r)^2$ & $(g - r)^3$ \\
\hline
$Z^{0}$ &  0  &  0  &  0  &  0 \\
$Z^{1}$ &  -0.962084  &  2.2796  &  4.16029  &  -3.27579 \\
$Z^{2}$ &  15.6602  &  -14.8073  &  19.261  &  4.28022 \\
$Z^{3}$ &  -82.9388  &  -49.2478  &  -40.9139  &  0 \\
$Z^{4}$ &  273.308  &  131.339  &  0  &  0 \\
$Z^{5}$ &  -312.677  &  0  &  0  &  0 \\
\hline
\end{tabular}
\end{table}

\begin{table}
\caption{Same as in Table~\ref{br07tabu_r} but for the SDSS $r$ band as a function of redshift and the $g - r$ colour.\label{br07tabrg_r}}
\begin{tabular}{l|cccc}
 & $(g - r)^0$ & $(g - r)^1$ & $(g - r)^2$ & $(g - r)^3$ \\
\hline
$Z^{0}$ &  0  &  0  &  0  &  0 \\
$Z^{1}$ &  -0.351251  &  2.61848  &  -2.99032  &  1.59058 \\
$Z^{2}$ &  1.93312  &  16.0682  &  -2.16736  &  -4.24709 \\
$Z^{3}$ &  -69.9339  &  -49.337  &  22.9267  &  0 \\
$Z^{4}$ &  253.373  &  12.0421  &  0  &  0 \\
$Z^{5}$ &  -235.32  &  0  &  0  &  0 \\
\hline
\end{tabular}
\end{table}

\begin{table}
\caption{Same as in Table~\ref{br07tabu_r} but for the SDSS $i$ band as a function of redshift and the $g - i$ colour.\label{br07tabg_i}}
\begin{tabular}{l|cccc}
 & $(g - i)^0$ & $(g - i)^1$ & $(g - i)^2$ & $(g - i)^3$ \\
\hline
$Z^{0}$ &  0  &  0  &  0  &  0 \\
$Z^{1}$ &  -5.58619  &  8.51897  &  -3.44266  &  0.823935 \\
$Z^{2}$ &  22.398  &  -31.8699  &  7.13812  &  -2.26748 \\
$Z^{3}$ &  -16.5911  &  40.9251  &  6.77963  &  0 \\
$Z^{4}$ &  -12.0117  &  -44.6466  &  0  &  0 \\
$Z^{5}$ &  21.0947  &  0  &  0  &  0 \\
\hline
\end{tabular}
\end{table}

\clearpage

\begin{table}
\caption{Same as in Table~\ref{br07tabu_r} but for the SDSS $z$ band as a function of redshift and the $r - z$ colour.\label{br07tabr_z}}
\begin{tabular}{l|cccc}
 & $(r - z)^0$ & $(r - z)^1$ & $(r - z)^2$ & $(r - z)^3$ \\
\hline
$Z^{0}$ &  0  &  0  &  0  &  0 \\
$Z^{1}$ &  -1.426  &  3.08833  &  -0.726039  &  1.06364 \\
$Z^{2}$ &  2.9386  &  -8.48028  &  -8.18852  &  -1.35281 \\
$Z^{3}$ &  8.08986  &  53.5534  &  13.6829  &  0 \\
$Z^{4}$ &  -93.2991  &  -77.1975  &  0  &  0 \\
$Z^{5}$ &  133.298  &  0  &  0  &  0 \\
\hline
\end{tabular}
\end{table}

\begin{table}
\caption{Same as in Table~\ref{br07tabu_r} but for the UKIDSS $Y$ band as a function of redshift and the $Y - H$ colour.\label{br07tabY_H}}
\begin{tabular}{l|cccc}
 & $(Y - H)^0$ & $(Y - H)^1$ & $(Y - H)^2$ & $(Y - H)^3$ \\
\hline
$Z^{0}$ &  0  &  0  &  0  &  0 \\
$Z^{1}$ &  -2.62137  &  4.7578  &  2.28856  &  -4.02782 \\
$Z^{2}$ &  29.4209  &  -26.5297  &  10.1083  &  10.5582 \\
$Z^{3}$ &  -141.372  &  34.7785  &  -41.2725  &  0 \\
$Z^{4}$ &  311.42  &  26.8742  &  0  &  0 \\
$Z^{5}$ &  -264.997  &  0  &  0  &  0 \\
\hline
\end{tabular}
\end{table}

\begin{table}
\caption{Same as in Table~\ref{br07tabu_r} but for the UKIDSS $J$ band as a function of redshift and the $J - K$ colour.\label{br07tabJJ_K}}
\begin{tabular}{l|cccc}
 & $(J - K)^0$ & $(J - K)^1$ & $(J - K)^2$ & $(J - K)^3$ \\
\hline
$Z^{0}$ &  0  &  0  &  0  &  0 \\
$Z^{1}$ &  -0.472236  &  2.1536  &  0.811858  &  -1.87211 \\
$Z^{2}$ &  -0.107502  &  -8.00546  &  16.6955  &  4.85177 \\
$Z^{3}$ &  -18.2002  &  -27.5709  &  -46.9334  &  0 \\
$Z^{4}$ &  111.89  &  99.1294  &  0  &  0 \\
$Z^{5}$ &  -162.057  &  0  &  0  &  0 \\
\hline
\end{tabular}
\end{table}

\begin{table}
\caption{Same as in Table~\ref{br07tabu_r} but for the UKIDSS $H$ band as a function of redshift and the $H - K$ colour.\label{br07tabH_K}}
\begin{tabular}{l|cccc}
 & $(H - K)^0$ & $(H - K)^1$ & $(H - K)^2$ & $(H - K)^3$ \\
\hline
$Z^{0}$ &  0  &  0  &  0  &  0 \\
$Z^{1}$ &  0.132484  &  1.93083  &  -4.7581  &  -6.67018 \\
$Z^{2}$ &  8.61784  &  -17.3496  &  28.7714  &  8.05742 \\
$Z^{3}$ &  -83.7003  &  267.725  &  127.193  &  109.678 \\
$Z^{4}$ &  134.398  &  -1657.94  &  -851.199  &  -217.543 \\
$Z^{5}$ &  567.048  &  4005.55  &  1071.17  &  0 \\
$Z^{6}$ &  -2000.79  &  -3298.45  &  0  &  0 \\
$Z^{7}$ &  1697.02  &  0  &  0  &  0 \\
\hline
\end{tabular}
\end{table}

\begin{table}
\caption{Same as in Table~\ref{br07tabu_r} but for the UKIDSS $K$ band as a function of redshift and the $J - K$ colour.\label{br07tabKJ_K}}
\begin{tabular}{l|cccc}
 & $(J - K)^0$ & $(J - K)^1$ & $(J - K)^2$ & $(J - K)^3$ \\
\hline
$Z^{0}$ &  0  &  0  &  0  &  0 \\
$Z^{1}$ &  -3.1771  &  1.66876  &  1.45967  &  -3.40684 \\
$Z^{2}$ &  17.9897  &  -7.83528  &  13.3436  &  9.32974 \\
$Z^{3}$ &  -114.067  &  -17.793  &  -42.0747  &  0 \\
$Z^{4}$ &  318.424  &  70.0829  &  0  &  0 \\
$Z^{5}$ &  -299.557  &  0  &  0  &  0 \\
\hline
\end{tabular}
\end{table}

\begin{table}
\caption{Same as in Table~\ref{tabu_r} but for the Johnson $U$ band as a function of
redshift and the $U - R_c$ colour.\label{tabU_R_c}}
\begin{tabular}{l|cccc}
 & $(U - R_c)^0$ & $(U - R_c)^1$ & $(U - R_c)^2$ & $(U - R_c)^3$ \\
\hline
$Z^{0}$ &  0  &  0  &  0  &  0 \\
$Z^{1}$ &  2.84791  &  2.31564  &  -0.411492  &  -0.0362256 \\
$Z^{2}$ &  -18.8238  &  13.2852  &  6.74212  &  -2.16222 \\
$Z^{3}$ &  -307.885  &  -124.303  &  -9.92117  &  12.7453 \\
$Z^{4}$ &  3040.57  &  428.811  &  -124.492  &  -14.3232 \\
$Z^{5}$ &  -10677.7  &  -39.2842  &  197.445  &  0 \\
$Z^{6}$ &  16022.4  &  -641.309  &  0  &  0 \\
$Z^{7}$ &  -8586.18  &  0  &  0  &  0 \\
\hline
\end{tabular}
\end{table}

\begin{table}
\caption{Same as in Table~\ref{tabu_r} but for the Johnson $B$ band as a function of
redshift and the $B - R_c$ colour.\label{tabB_R_c}}
\begin{tabular}{l|cccc}
 & $(B - R_c)^0$ & $(B - R_c)^1$ & $(B - R_c)^2$ & $(B - R_c)^3$ \\
\hline
$Z^{0}$ &  0  &  0  &  0  &  0 \\
$Z^{1}$ &  -1.99412  &  3.45377  &  0.818214  &  -0.630543 \\
$Z^{2}$ &  15.9592  &  -3.99873  &  6.44175  &  0.828667 \\
$Z^{3}$ &  -101.876  &  -44.4243  &  -12.6224  &  0 \\
$Z^{4}$ &  299.29  &  86.789  &  0  &  0 \\
$Z^{5}$ &  -304.526  &  0  &  0  &  0 \\
\hline
\end{tabular}
\end{table}

\begin{table}
\caption{Same as in Table~\ref{tabu_r} but for the Johnson $V$ band as a function of
redshift and the $V - I_c$ colour.\label{tabV_I_c}}
\begin{tabular}{l|cccc}
 & $(V - I_c)^0$ & $(V - I_c)^1$ & $(V - I_c)^2$ & $(V - I_c)^3$ \\
\hline
$Z^{0}$ &  0  &  0  &  0  &  0 \\
$Z^{1}$ &  -1.37734  &  -1.3982  &  4.76093  &  -1.59598 \\
$Z^{2}$ &  19.0533  &  -17.9194  &  8.32856  &  0.622176 \\
$Z^{3}$ &  -86.9899  &  -13.6809  &  -9.25747  &  0 \\
$Z^{4}$ &  305.09  &  39.4246  &  0  &  0 \\
$Z^{5}$ &  -324.357  &  0  &  0  &  0 \\
\hline
\end{tabular}
\end{table}

\begin{table}
\caption{Same as in Table~\ref{tabu_r} but for the Cousins $R_c$ band as a function of
redshift and the $B - R_c$ colour.\label{tabR_c_B_R_c}}
\begin{tabular}{l|cccc}
 & $(B - R_c)^0$ & $(B - R_c)^1$ & $(B - R_c)^2$ & $(B - R_c)^3$ \\
\hline
$Z^{0}$ &  0  &  0  &  0  &  0 \\
$Z^{1}$ &  -2.83216  &  4.64989  &  -2.86494  &  0.90422 \\
$Z^{2}$ &  4.97464  &  5.34587  &  0.408024  &  -2.47204 \\
$Z^{3}$ &  -57.3361  &  -30.3302  &  18.4741  &  0 \\
$Z^{4}$ &  224.219  &  -19.3575  &  0  &  0 \\
$Z^{5}$ &  -194.829  &  0  &  0  &  0 \\
\hline
\end{tabular}
\end{table}

\begin{table}
\caption{Same as in Table~\ref{tabu_r} but for the Cousins $I_c$ band as a function of
redshift and the $V - I_c$ colour.\label{tabI_c_V_I_c}}
\begin{tabular}{l|cccc}
 & $(V - I_c)^0$ & $(V - I_c)^1$ & $(V - I_c)^2$ & $(V - I_c)^3$ \\
\hline
$Z^{0}$ &  0  &  0  &  0  &  0 \\
$Z^{1}$ &  -7.92467  &  17.6389  &  -15.2414  &  5.12562 \\
$Z^{2}$ &  15.7555  &  -1.99263  &  10.663  &  -10.8329 \\
$Z^{3}$ &  -88.0145  &  -42.9575  &  46.7401  &  0 \\
$Z^{4}$ &  266.377  &  -67.5785  &  0  &  0 \\
$Z^{5}$ &  -164.217  &  0  &  0  &  0 \\
\hline
\end{tabular}
\end{table}

\section{K-corrections calculator}

To facilitate the usage of described analytical approximations and
calculation of necessary $K$-corrections for user's data, we provide a
dedicated web-site entitled ``K-corrections
calculator''\footnote{http://kcor.sai.msu.ru/} offering several ways
of operation. Firstly, one can determine $K$-corrections for objects
of interest one by one by manually filling interactive web form with
the data available. This is the most straightforward method to
determine $K$-corrections and it works in any modern web browser. As a
second option, the ready-to-use code snippets written in popular data
languages (C, IDL, Python) together with the corresponding
analytical expressions are provided to simplify the integration of
$K$-correction functionality into the user's code. Both ways
allow a user to calculate $K$-corrections for SDSS $ugriz$ and UKIRT
WFCAM $YJHK$ (in AB magnitudes), but also for Johnson-Cousins $UBVR_cI_c$ and 2MASS $JHK_s$
bands (in Vega magnitudes), choosing the most convenient colour as an input argument from
several options. Based on user demands and feedback, we will consider
the development of a web service for bulk calculation of
$K$-corrections and publication of additional filter-colour
combinations in the calculator.

\label{lastpage}

\end{document}